\pdfoutput=1
\documentclass[12pt]{emulateapj}

\newcommand{\htwo}{H{\footnotesize II }}
\newcommand{\brem}{{\it bremsstrahlung }}

\newcommand{\Msun}{\mbox{$M_{\odot}$}}

\slugcomment{}

\shorttitle{MASSIVE CLUMPS IN NGC 6334}
\shortauthors{Mu\~noz et al.}

\begin{document}

\title{MASSIVE CLUMPS IN THE NGC~6334 STAR FORMING REGION}
\bigskip
\bigskip
\author{Diego J. Mu\~noz\altaffilmark{1}, Diego Mardones,  Guido Garay and David Rebolledo}
\affil{Departamento de Astronom\'ia, Universidad de Chile,
    Casilla 36-D, Santiago, Chile}
\author{Kate Brooks}
\affil{Australia Telescope National Facility,P.O. Box 76,
Epping NSW 1710 Australia}
\smallskip
\and
\author{Sylvain Bontemps}
\affil{Observatoire de Bordeaux, 2 rue de l'Observatoire,33270 Floirac, France}

\altaffiltext{1}{dmunoz@cfa.harvard.edu}

\begin{abstract}
We report observations of dust continuum emission at 1.2 mm towards 
the star forming region NGC 6334 made with the SEST SIMBA bolometer array.
The observations cover an area of $\sim 2$ square degrees with
approximately uniform noise.  We detected 181 clumps spanning almost
three orders of magnitude in mass (3\Msun$-6\times10^3$ \Msun) and with sizes in 
the range 0.1--1.0~pc.  We find that the clump mass function 
$dN/d\log M$ is well fit with a power law of the mass with exponent
$-0.6$ (or equivalently $dN/dM \propto M^{-1.6}$). The derived exponent  is 
similar to those obtained from molecular line emission surveys and is 
significantly different from that of the stellar initial mass function.  
We investigated changes in the mass spectrum by changing the assumptions
on the temperature distribution of the clumps and on the contribution of 
free-free emission to the 1.2 mm emission, and found little changes 
on the exponent. The Cumulative Mass Distribution Function is also 
analyzed giving consistent results in a mass range excluding the 
high-mass end where a power-law fit is no longer valid. 
The masses and sizes of the clumps observed in NGC 6334 indicate that 
they are not direct progenitors of stars and that the process of 
fragmentation determines the distribution of masses later on or occurs 
at smaller spatial scales.

The spatial distribution of the clumps in NGC 6334 reveals clustering 
which is strikingly similar to that exhibited by young stars in other 
star forming regions. A power law fit to the surface density of companions 
gives $\Sigma\propto \theta^{-0.62}$. 

\end{abstract}


\section{INTRODUCTION}
\subsection{Massive Star Formation and Molecular Cloud Structure}

Half of the mass in the interstellar medium is in the form of
molecular gas exhibiting a broad range of structures, ranging
from small isolated clouds, with masses of a few $M_\odot$ and subparsec
sizes, to Giant Molecular Clouds (GMCs), with masses of
several times $10^6 M_\odot$ and sizes of 100~pc \citep{bli93}. 
We adopt the nomenclature used in the review of \citet{will00}
to refer to the different observed molecular structures.
GMCs are the sites of most of star formation activity in the Milky Way,
and in particular of high mass stars,  which are usually born in
clusters within massive cores. In order to understand how 
massive cores form from the GMC complexes we must understand 
how fragmentation and condensation proceed within them.
It is essential for this purpose to determine the physical properties 
of complete samples of massive cores within GMCs.

Molecular line surveys, at millimeter and sub-millimeter wavelengths, have 
revealed the structure of GMCs to be highly inhomogeneous and clumpy
\citep{bli93,eva99}. These surveys
have shown that the mass spectra of clouds \citep{san85,solo87},
clumps \citep{bli93,kram98,elad99,will00} and total mass of embedded clusters \citep{lad03} 
are similar to one another. These mass spectra are 
notably different than the stellar mass spectrum: the Initial Mass Function (IMF;  \citealp*{sal55}. In particular,
several works on molecular line mapping of GMCs, 
show that their mass spectra follow a power law with nearly the 
same exponent, $x\sim 0.6$, where $dN/d\log M \propto M^{-x}$
(e.g. \citealp*{bli93,will00}). 

The development of large bolometer arrays during the last ten years
has permitted to carry out extended millimeter and sub-millimeter continuum surveys, 
allowing the direct dust mass determination of clumps.
Motte et al (1998) mapped the Ophiuchus cloud with the IRAM 30-m telescope and found a distribution of clump masses similar to Salpeter's IMF.
Several works \citep{john01,beu04,mook05,rei05,john06}
have reported mass spectra of dust cores with indexes similar to Salpeter's 
IMF, and different from those derived from molecular line studies.
If the mass function of cores is similar to the Salpeter IMF,
independent of the range of masses involved,  then the
star formation process within GMCs would be defined 
in the earliest stages as a result of cloud fragmentation.
\newpage
In this paper we present a large scale 1.2~mm continuum study
of the NGC~6334 GMC aimed to find and study a complete sample of massive cores.

\begin{center}
\begin{figure*}
\centering
\plotone{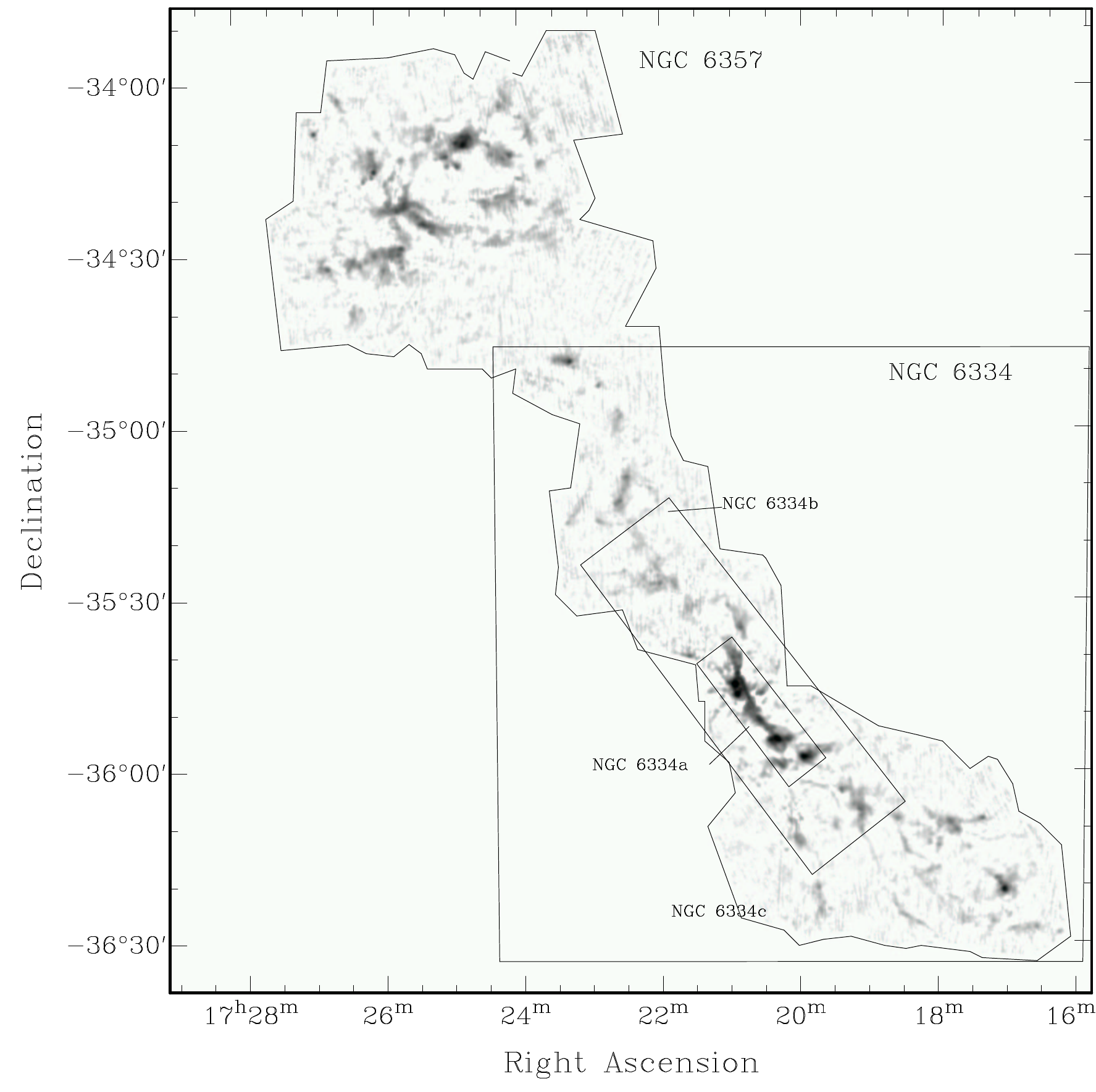}
\caption{
Grey scale image of the 1.2~mm emission towards NGC 6334
and NGC 6357.  The polygonal line was used to exclude the noisy borders of the mapped region.
The large square indicates the area defined 
as NGC 6334 in the present analysis. It encloses 182 of the
347 clumps found using {\it clfind2d}.  Also labeled are the three
sub-regions chosen for statistical analysis: the central region NGC 6334a (see
Figure~\ref{radio_sources_2}); a larger extension NGC~6334b; and finally 
NGC ~6334c which includes only the clumps  outside of NGC 6334b.
\label{final2_region}}
\end{figure*}
\end{center}

\subsection{NGC~6334 and NGC~6357}

NGC~6334 is one of the nearest and most prominent sites of massive
star formation,  at a distance of only 1.7 kpc \citep{neck78}. The central
region of NGC~6334 consists of a $\sim10$~pc long filament with
seven sites of massive star formation.
Within them there is a wide variety of activity associated with star
formation, such as water masers, \htwo regions \citep{rod82,carr02}, and
molecular outflows.

Early FIR \citep{mcbree79} and radio \citep{rod82} surveys characterized
the overall properties of the main sites of massive star formation, 
whereas NIR studies \citep{stra89a,stra89b} revealed the
cluster forming nature of many of these sites. 
\cite{krae99b} and \cite{burt00} review the different
notations used by previous authors to identify bright sources.
Sub-millimetric and millimetric results on NGC~6334 have been 
previously reported by \citet{gez82}, \citet{san00} and \citet{mccut00}, 
all of which focused in the northern portion of the main filament:
sources I and I(N) which have been recently resolved into smaller cores
\citep{hun06}. In the case of \citet{san00}, the sources I and I(N) were
redefined as the central peaks of each respective source. This after 
observations at 350~$\mu$m, 450~$\mu$m, 380~$\mu$m, 1.1~mm and 1.3~mm
showed inner structure for Gezari's cores. In \citet{san00} sources I and I(N) have angular sizes of
$10''\times8''$ and $11''\times8''$ respectively. These sizes are $\sim10$ times smaller than our massive
clumps cl1 and cl2 identified as I and I(N) respectively.
For the present work, cl1 encloses sources I and I(NW) of Sandell's nomenclature, while cl2 encloses
I(N), SM1,SM2,SM4 and SM5 as well as considerable extended emission in both cases. Hence, the
comparison between that work and ours is not straighforward.

Following the evolutionary sequence proposed by \citet{beu06},
NGC~6334 and NGC~6357 appear to be in an intermediate phase of Massive
Starless Clumps and Protoclusters. The latter region seems more
evolved that the former since compact \htwo regions are still
prominent in NGC~6334 at radio wavelengths while the structure of
NGC~6357 is more disrupted, suggesting that massive stars have already shaped the mother clouds. This is also supported by the presence of more infrared sources
in NGC~6357.   From this evidence, and the lack of significant
amounts of cold material in between the two regions,  
we will consider NGC~6334 and
NGC~6357 as two independent regions, defining NGC~6334
as the southeastern portion of the map as shown in
Figure~\ref{final2_region}.

Even though NGC~6334 is one of the closest GMCs, it is still far
compared to low-mass star-forming regions (e.g.\ Ophiucus, Orion B,
Taurus). As a consequence, the detected clumps in NGC~6334 are larger and
considerably more massive, and can be considered likely cluster-forming cores
\citep{mott03,war06,beu06} and we are unable to resolve their inner
structure.

\section{Observations and Data Reduction}

The regions NGC~6334 and NGC~6357 were mapped using the 37 channel SEST
Imaging Bolometer Array (SIMBA) in the fast mapping mode in three
different epochs: July 2002, September 2002 and May 2003. The
passband of the bolometer has an equivalent width of 90 GHz and
is centered at 250 GHz (1.2~mm).
The half-power beamwidth of the instrument is 24\arcsec\ giving a
spatial resolution of ~0.2 pc. Ninety five observing blocks were
taken towards the NGC~6334 and NGC~6357 regions, with typical extension
of $\sim10'\times20'$, to sample a total area of $\sim 2$ square
degrees between $17^h16^m00^s\;-36^\circ40'00''$ and
$17^h28^m00^s\;-33^\circ40'00''$.  Skydip observations were done
approximately every two hours to determine the zenith opacity at
250 GHz.  Typical opacities were $\tau\approx 0.2$ with values
ranging from 0.17 to 0.4 in a few cases. We also checked pointing
on $\eta$~Carinae every two hours and found a typical rms
deviation of 3-5'' in azimuth and elevation. Every night we
observed Uranus for flux calibration.

The SIMBA data were reduced using the MOPSI program written by
Robert Zylka after conversion by the simbaread program written at
ESO. The SIMBA raw data consists of a time series for each of the
37 bolometers (channels) in the array.    The time series includes
the counts per channel and sky position. The reduction procedure
first removes the brightest data spikes. Next a low order baseline
in time is fit to the full observation file for each channel, and
a zero order baseline is fit in azimuth for each channel.  The
data is then deconvolved by the time response function of each
channel, as measured by the SEST staff.    Gain elevation and
extinction corrections are applied next.    An iterative  sky
noise reduction procedure is then applied, where the counts of
each channel are correlated with those of the other 36 channels,
yielding a so-called flat field correction to calibrate the
relative sensitivity of each pixel.   A source image is finally
produced by averaging the flux of all channels as they pass
through the same position on the sky.

The sky noise reduction algorithm includes the flux coming from
both the source and the sky simultaneously.  If the source extends
over several channels, this introduces spurious correlated flux
which hampers the sky noise reduction procedure. To avoid this, a
smoothed model of the source flux distribution on the sky is subtracted 
from the raw time series data. Thus, the sky noise
reduction procedure can be repeated, finding a better source model
with each iteration. 

The calibration was derived from maps of Uranus. The resulting
multiplicative factor varied between 0.06-0.09
Jy~count$^{-1}$~beam$^{-1}$. Finally, reduced images were combined
with the MOPSI software to produce the final map (Figure~\ref{final2_region}). 
In order to reduce noise further, a Gaussian smoothing of
30'' was applied to the final image and the map edges were
removed. The typical rms noise of the final map is 25
mJy~beam$^{-1}$.  SIMBA observations usually  have an absolute flux
uncertainty of $\sim$20\% \citep{fau04}.

\section{RESULTS}
 \label{sec:results}

\subsection{Clump-finding Algorithms}

\begin{figure*}
\plottwo{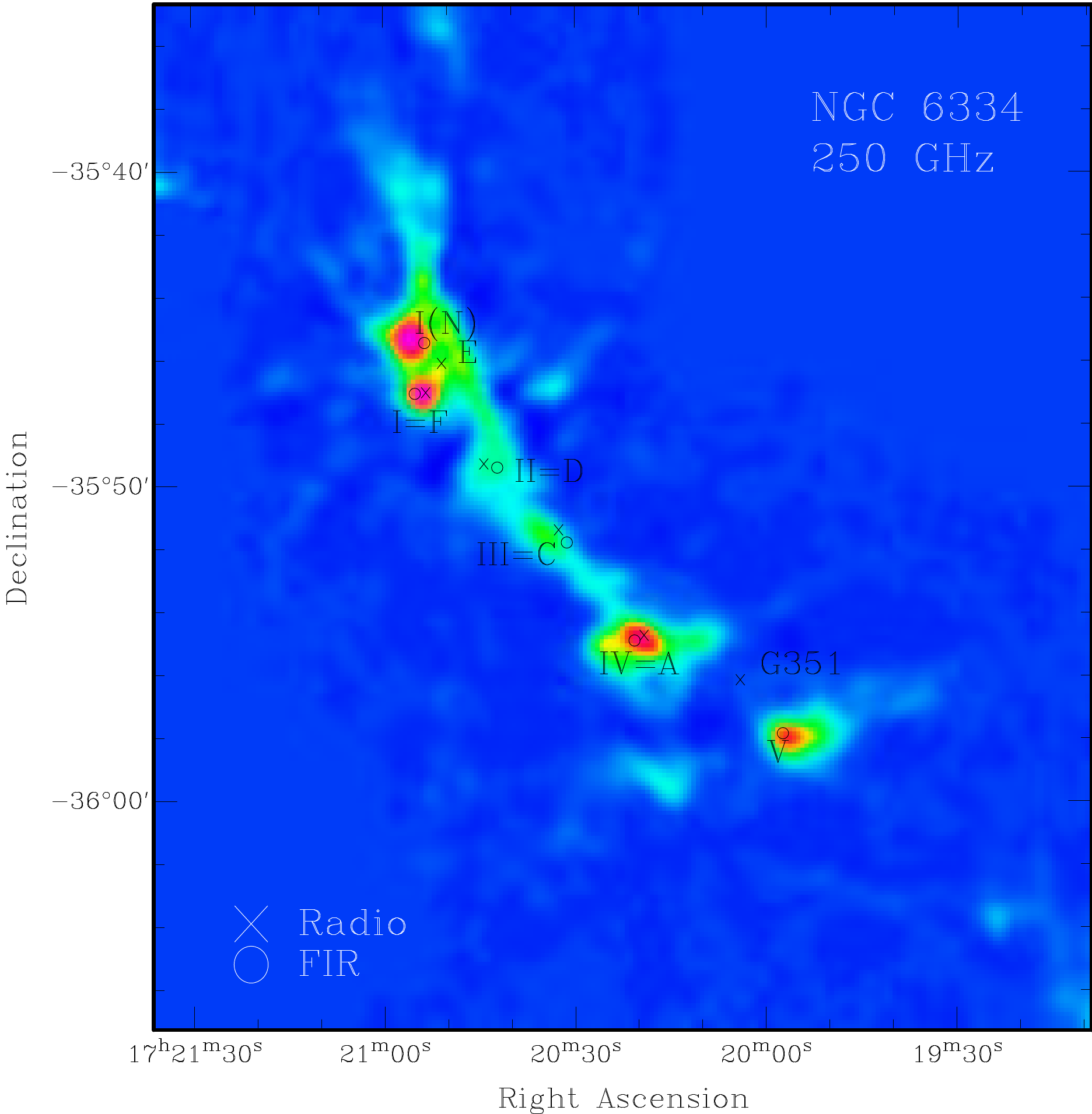}{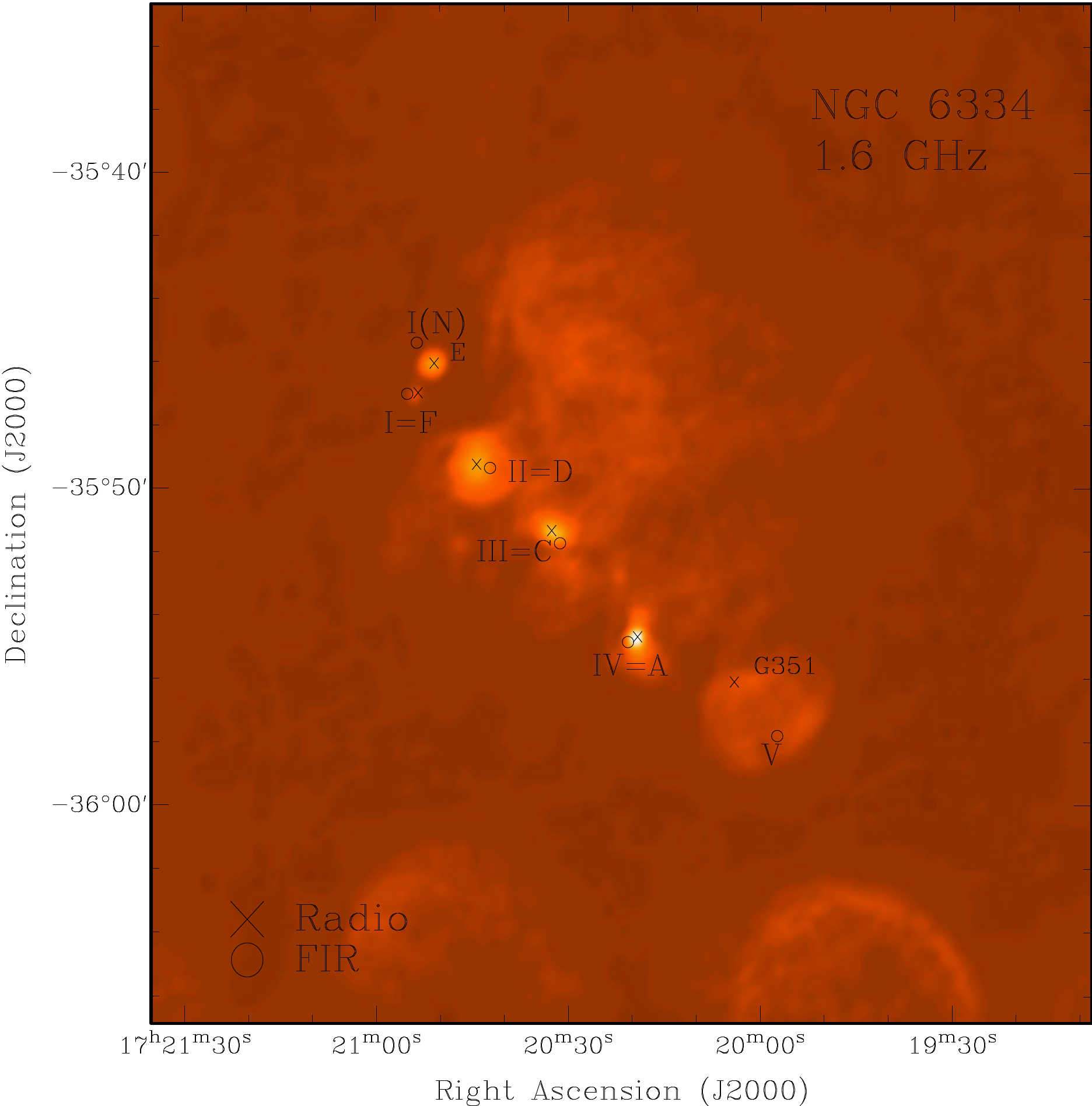}
\caption{ Peak position of the radio and FIR sources of
Table~\ref{tbl-sources} plotted over images of the SIMBA continuum at 
250~GHz (left panel) and ATCA continuum at 1.6~GHz (right panel). 
\label{radio_sources_2}}
\end{figure*}

Different clump-finding algorithms have been used to study the
substructure in molecular clouds.
Among the most used are {\it Clumpfind}
\citep{will94}  and {\it Gaussclumps} \citep{stut90}.   Both algorithms have different biases, but find similar clump distributions (e.g. \citealp*{schnei04}).
We use the clumpfind algorithm because it makes no
assumptions about inherent clump shapes. Clumpfind first finds the brightest emission peak in the image, then it descends to a lower contour level and finds
all the image pixels above this level, associating them to the first peak
(clump) if contiguous, or else defines one or more additional clumps.
The spatial separation
of clumps is defined along saddle points.  We used a conservative lower detection threshold 
of 75 mJy ($3\sigma$) per beam. 
We found 347 clumps in 
the whole image, 182 of which are in the NGC~6334 region.  Only two clumps are 
likely to be fictitious based on their small effective radii and location 
close to the borders, and only one of them is in the NGC6334 region. Thus, 
we will use the remaining 181 clumps in our analysis of the region NGC~6334 
in this paper.
Table \ref{tbl-sources} lists the clumps in our sample 
associated with previously known radio and IR sources. 
These sources are all located in the brightest part
of the filament as seen in Figure~\ref{radio_sources_2}.

\begin{deluxetable*}{lllllll}
\tabletypesize{\scriptsize} \tablecaption{Radio, FIR and
Millimeter Sources in NGC~6334 \label{tbl-sources}}
 \tablewidth{0pt}
\tablehead{ \colhead{Radio} & \colhead{Radio Position} &
\colhead{FIR} & \colhead{FIR Position} & \colhead{SIMBA}
& \colhead{SIMBA Position} &\colhead{Ref} \\
\colhead{Name}&\colhead{}&\colhead{Name}&\colhead{}&\colhead{Name}&\colhead{}& \colhead{}} 
\startdata 
&$\;\;\;\;\;\;\;\;\alpha\;\;\;\;\;\;\;\;\;\;
\;\;\;\;\;\;\;\;\;\;\delta$&&   $\;\;\;\;\;\;\;\;\alpha\;\;\;\;\;
\;\;\;\;\;\;\;\;\;\;\;\;\;\;\;\delta$&&
$\;\;\;\;\;\;\;\;\alpha\;\;\;\;\;
\;\;\;\;\;\;\;\;\;\;\;\;\;\;\;\delta$\\
& $\;\;\;\;\;\;\;\;\;\;\;\;\;\;$ (J2000)&  & $\;\;\;\;\;\;\;\;\;\;\;\;\;\;$(J2000)&&$\;\;\;\;\;\;\;\;\;\;\;\;\;\;\;$(J2000)\\
\tableline

G351.20+0.70& 17$^h$20$^m$04.1$^s\;\;\;$-35$^\mathrm{\circ}$56'10''  &&&cl210& 17$^h$19$^m$58.0$^s\;\;\;$-35$^\mathrm{\circ}$55'56'' &(4),$\dag$\\

unnamed\tablenotemark{a}&&V&  17$^h$19$^m$57.4$^s\;\;\;$-35$^\mathrm{\circ}$57'52''  &cl4& 17$^h$19$^m$56.7$^s\;\;\;$-35$^\mathrm{\circ}$57'56''& (2),$\dag$\\

A&17$^h$20$^m$19.2$^s\;\;\;$-35$^\mathrm{\circ}$54'45'' &IV& 17$^h$20$^m$20.7$^s\;\;\;$-35$^\mathrm{\circ}$54'55''  &cl3 (cl7,cl15)& 17$^h$20$^m$19.8$^s\;\;\;$-35$^\mathrm{\circ}$54'44''& (1),(2),$\dag$ \\

C&17$^h$20$^m$32.6$^s\;\;\;$-35$^\mathrm{\circ}$51'24'' &III& 17$^h$20$^m$31.3$^s\;\;\;$-35$^\mathrm{\circ}$51'48'' &cl11& 17$^h$20$^m$34.2$^s\;\;\;$-35$^\mathrm{\circ}$51'32'' & (1),(2),$\dag$\\
D&17$^h$20$^m$44.3$^s\;\;\;$-35$^\mathrm{\circ}$49'18'' &II& 17$^h$20$^m$42.2$^s\;\;\;$-35$^\mathrm{\circ}$49'25'' &cl19& 17$^h$20$^m$42.8$^s\;\;\;$-35$^\mathrm{\circ}$49'16''& (1),(2),$\dag$\\

E&17$^h$20$^m$50.9$^s\;\;\;$-35$^\mathrm{\circ}$46'06'' &\nodata&\nodata&\nodata&\nodata &(1)\\

F&17$^h$20$^m$53.4$^s\;\;\;$-35$^\mathrm{\circ}$47'02'' &I& 17$^h$20$^m$55.1$^s\;\;\;$-35$^\mathrm{\circ}$47'04'' &cl1& 17$^h$20$^m$54.0$^s\;\;\;$-35$^\mathrm{\circ}$47' 00'' & (1),(2),$\dag$\\
\nodata&\nodata&I(N)&
17$^h$20$^m$53.6$^s\;\;\;$-35$^\mathrm{\circ}$45'27'' &cl2&
17$^h$20$^m$56.0$^s\;\;\;$-35$^\mathrm{\circ}$45'16''&
(3),$\dag$
\enddata

\tablecomments{The FIR and Radio Equatorial coordinates have
been precessed from B1950 to J2000.}
 \tablenotetext{a}{The radio counterpart for FIR source V is observed at
1.6 Ghz with
similar peak intensity to G351.20+0.70, but was not identified earlier as
 an HII region
but as part of a PDR shell \citep{mor90,jack99,burt00}} \tablerefs{
(1)\citet{rod82},(2)\citet{mcbree79},
 (3)\citet{gez82}, (4)\citet{mor90},
 $\dag$ This work }
\end{deluxetable*}

\subsection{Clump Size Distribution}

\begin{figure}[h]
\plotone{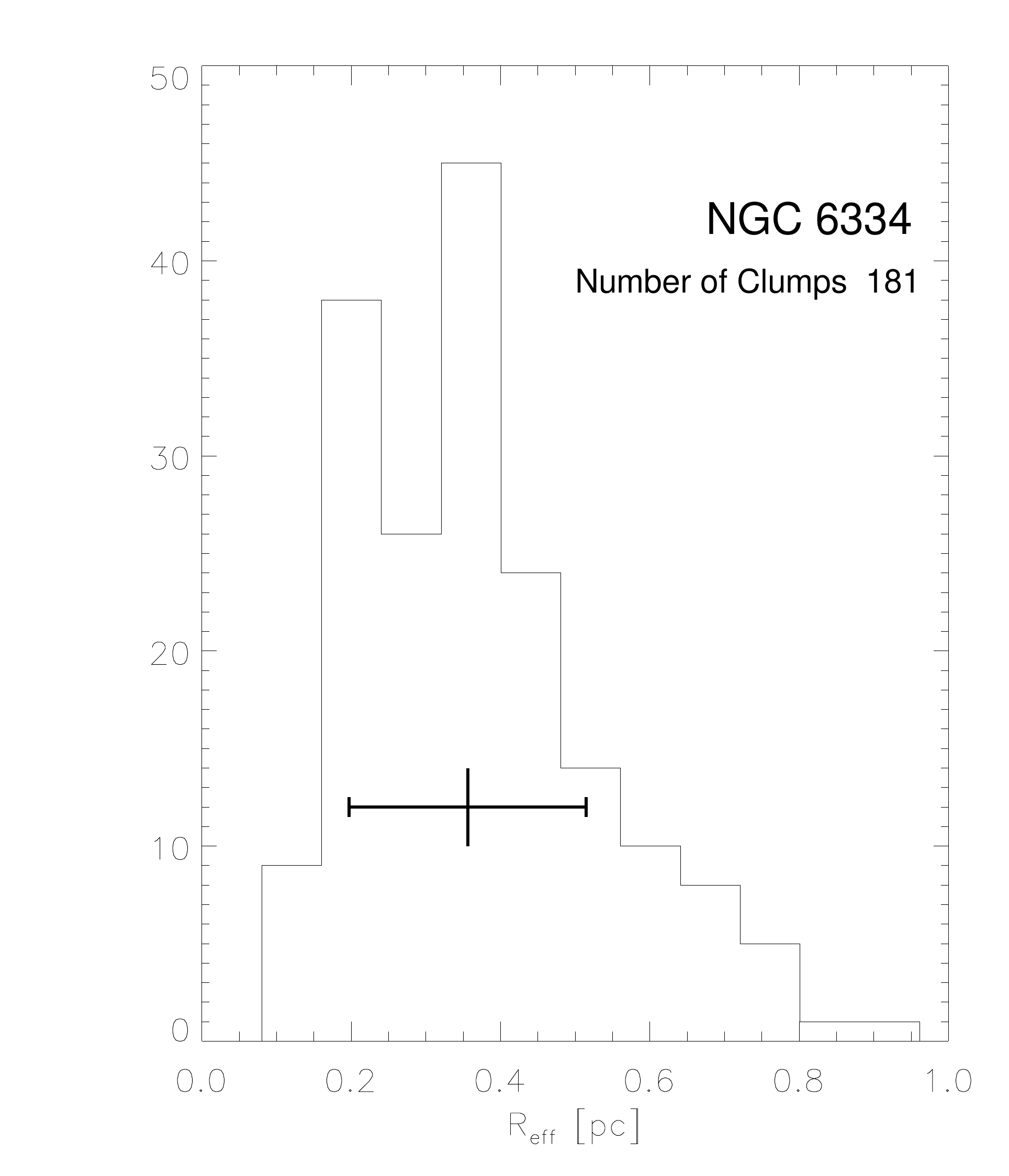}
\caption{Clump Size Distribution (in pc) assuming a distance
of 1.7~kpc to NGC~6334.  
The cross below the histogram indicates the median and standard deviation of the distribution.  
\label{clump_sizes}}
\end{figure}

Figure~\ref{clump_sizes} shows a histogram of the clump size distribution.
The effective radius, or size, is determined from the angular area encompassed 
by each clump (from the {\it clfind2d} output) assuming a distance
to NGC~6334 of 1.7~kpc. The sizes range from 0.1 to 1.0~pc, with a median of 
0.36~pc. These values are similar to those derived for 
clumps within GMCs \citep{bli93,will00,pud02,fau04,beu06}.

\subsection{Mass Estimates}

Since the dust emission at 1.2 mm is most likely to be optically thin 
(e.g.\ Garay et al.\ 2002), the mass of each clump can be estimated from the observed 
flux density. For an isothermal dust source, the total gas mass $M_g$ is 
related to the observed
flux density $S_\nu$ at an optically thin frequency $\nu$ as (c.f. \citealp*{chi87})
\begin{equation}\label{eq:mass}
M_g=\frac{S_\nu D^2}{R_{dg}\kappa_\nu B_\nu(T_d)},
\end{equation}
where $\kappa_\nu$ is the dust mass absorption coefficient, $D$ is the source
distance, $R_{dg}$ is the dust to gas mass ratio, and $B_{\nu}$ is the
Planck function.     In more convenient units the gas mass can be written as
\begin{eqnarray}\label{eq:mass2}
\nonumber M_g=20.4\left(\frac{S_\mathrm{250GHz}}{\mathrm{Jy}} \right)
\left(\frac{D}{\mathrm{kpc}}  \right)^2
\left( \frac{0.01}{R_{dg}} \right) \\
\times  \left( \frac{1\mathrm{cm}^2\mathrm{g}^{-1}}
{\kappa_\mathrm{250GHz}} \right) \left[e^\frac{12\,\mathrm{K}}{T_d}-1
\right]M_\odot.
\end{eqnarray}

Using a dust mass absorption coefficient of
$\kappa_\mathrm{250~GHz}=1~\mathrm{cm}^2\mathrm{g}^{-1}$
\citep{oss94}; a dust-to-gas ratio of 0.01; a distance 
of 1.7~kpc, and a dust temperature $T_d=17$~K \footnote{
We choose a value of 17~K because it lies in the typical range for cold dark clouds \citep{pud02}.
 It is also
also a factor of two smaller 34~K 
Ð the average temperature of clouds with infrared counterparts  \citep{fau04}Ð
making the comparison between both temperatures easier. 
}, we
computed the mass of each clump found by {\it clfind2d}.
We find that the clump 
masses in NGC~6334 range from 3 to $6000\,M_\odot$, with a mean value 
of $170\,M_\odot$ and a median value of $60\,M_\odot$.
The total mass of the clumps in NGC~6334 
is $M_\mathrm{tot}\sim5\times10^4\,M_\odot$.

\subsection{Clump Mass Spectrum}
\label{sec:spectrum}

The Clump Mass Function (CMF), $\xi_{\alpha}(M)$, is defined as the number 
of clumps per unit mass,
\begin{equation}
  \xi_{\alpha}(M) = \frac{dN}{dM}  \approx \frac{\Delta N}{\Delta M} ,
\end{equation}
where $\Delta N$ and $\Delta M$ are used to indicate observational estimates. 
Some workers prefer to use a {\sl logarithmic} CMF, $\xi_x$, defined as the 
number of clumps per unit logarithmic mass,
\begin{equation}
  \xi_{x}(M) = \frac{dN}{d(\log M)} \approx \frac{\Delta N}{\Delta (\log M)} .
\end{equation}
These two functions are related by the expression $ \xi_x(M) = (M ln10)\xi_{\alpha}(M)$.   
If $\xi_{\alpha}(M)$ has a power law dependence with mass, 
$\xi_{\alpha}(M)\propto M^{-\alpha}$, 
then the logarithmic CMF should also have a power law 
dependence with mass, $\xi_{x}(M)\propto M^{-x}$, 
where the power law exponents are related by $x = \alpha -1$.
A Salpeter slope corresponds to $x=1.35$ or $\alpha = 2.35$.

\begin{figure}
\plotone{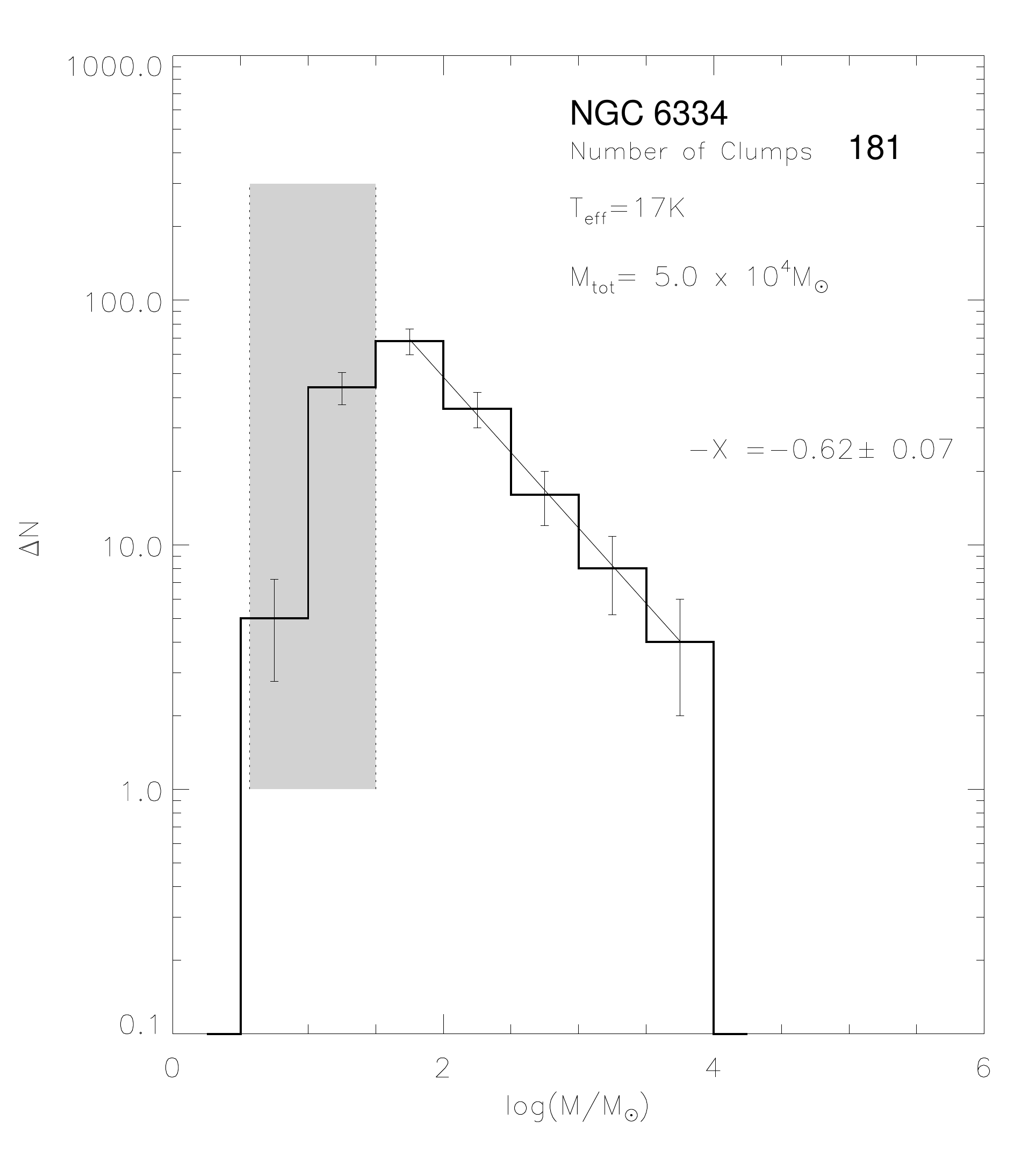}
 \caption{Mass histogram of the clumps within NGC~6334.  
 $\Delta N$ is the number of clumps in bins of constant $\Delta\log M$ (=0.5).
The shaded region delimits the $3\sigma$ detection limit and our estimated 
completeness limit. This region is excluded from the least squares fit. 
 \label{6334histo_1}}
\end{figure}

The CMF is usually estimated from a histogram of the derived clump masses 
assuming it has a power law form. The distinction between the observationally 
derived $\xi_{\alpha}$ and $\xi_{x}$ is made by \citet{sca98}, \citet{kro01} 
and \citet{lar03}.  Its implications in the type of binning used in the 
histogram is also mentioned in \citet{kle00}.
Figure \ref{6334histo_1} shows an histogram of the mass of the clumps within 
NGC~6334.   The bin size $\Delta \log M$ has a constant value of 0.5. 
The completeness limit is estimated to be $\sim 30\,M_\odot$.
A least squares linear fit to the $\Delta N$ versus 
$\log M$ relationship gives a slope of $-0.62\pm0.07$.

\begin{figure*}[h]
\epsscale{0.3}
\plotone{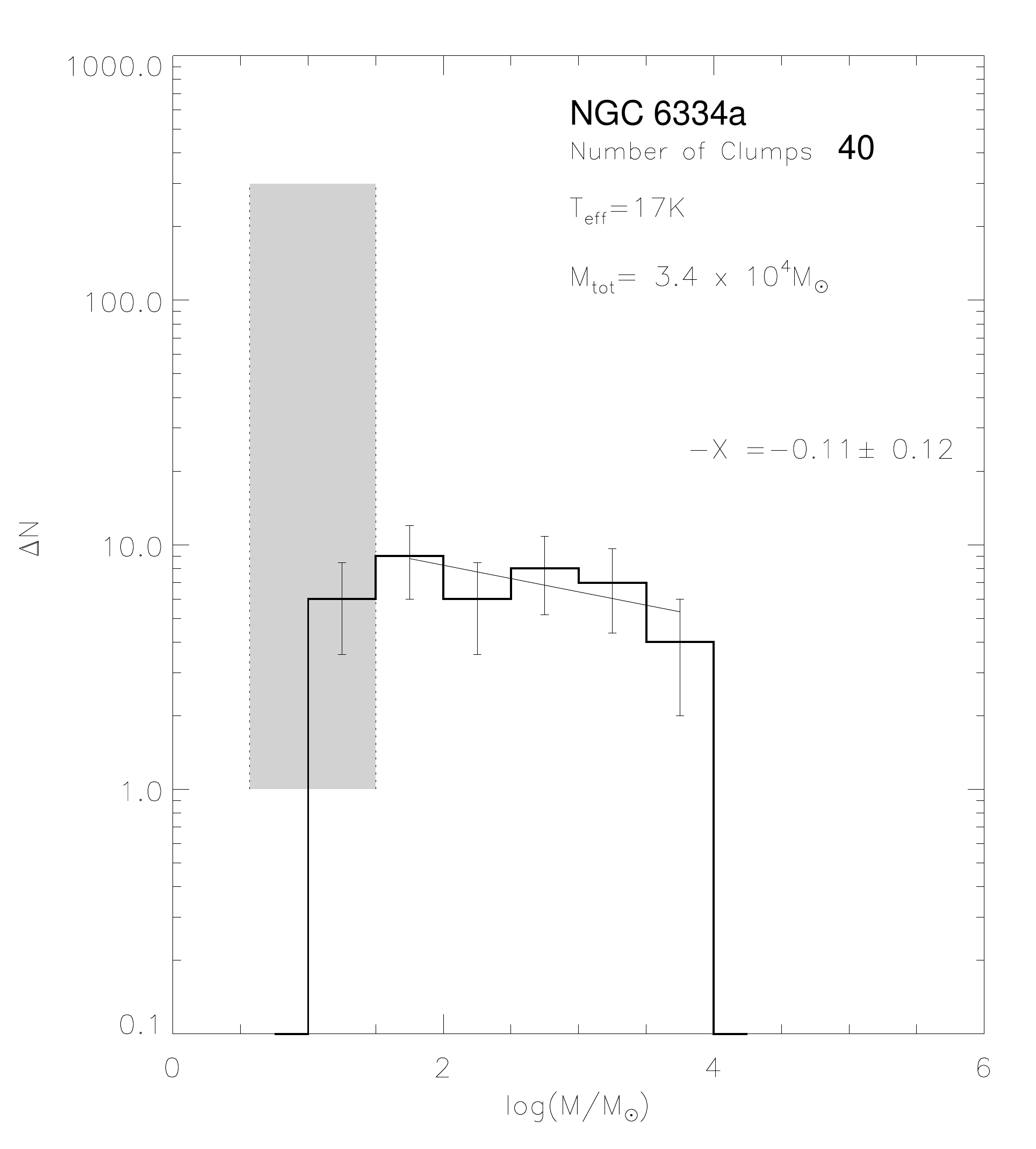}
\plotone{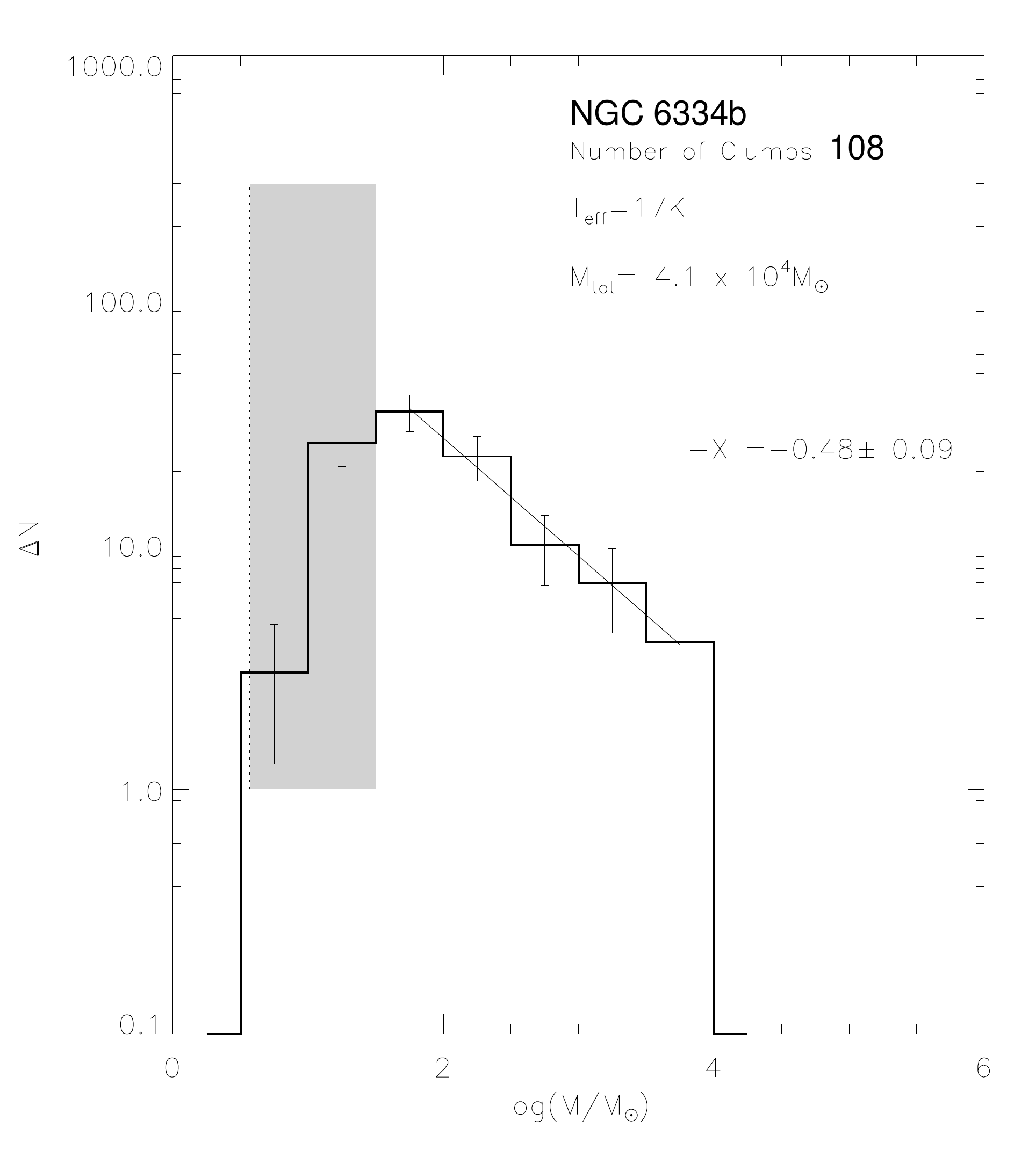}
\plotone{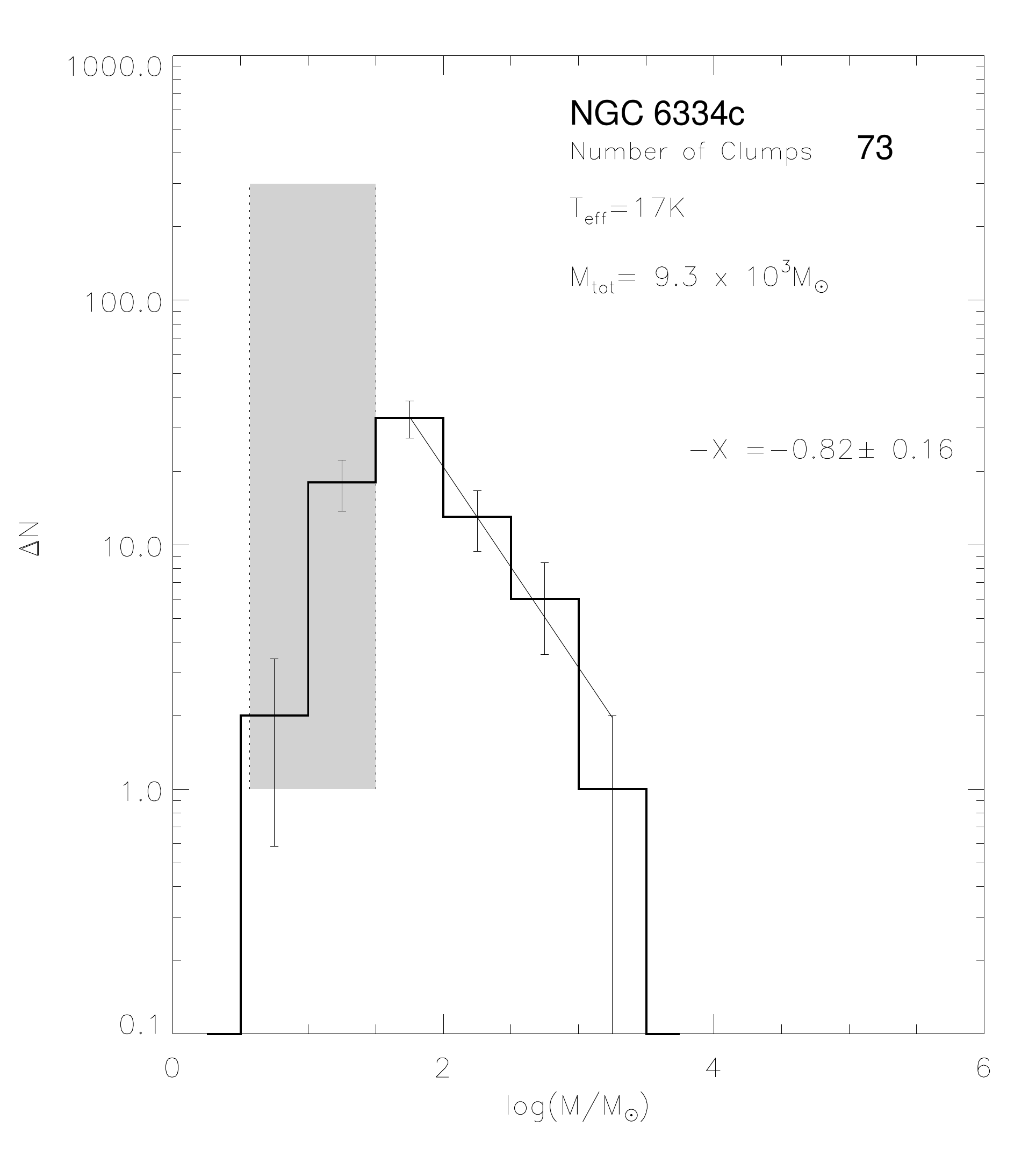}
\epsscale{1.0}
\caption{Mass histogram of clumps within selected regions of NGC~6334
 (see Figure~\ref{final2_region}). 
 Top panel: NGC~6334a. $x=0.11$
 Middle panel: NGC~6334b. $x=0.48$
 Bottom panel: NGC~6334c. $x=0.82$ 
\label{histograms_abc}}
\end{figure*}

\paragraph{Variations of the CMF within NGC~6334.}
Given the large number of clumps in our survey, we can 
assess possible changes in the CMF as a function of position within the cloud.
We consider the three subregions within NGC~6334 defined
in Figure~\ref{final2_region} and construct a mass spectrum for each 
of them (see Figure~\ref{histograms_abc}). The CMF in NGC~6334a,
which encompasses the central filament containing the most massive clumps 
and where star formation is clearly taking place, is well fitted with a power law
with an exponent of $x=0.11\pm0.12$, significantly shallower than the value 
determined for the whole sample.   The exponent steepens for NGC~6334b ($x=0.48$), which covers a much larger region than the central filament.  For NGC~6334c, which excludes the
main filament, the slope is even steeper, $x=0.82$.  Thus, the slope of the CMF in NGC~6334 depends on the location of the clumps within the cloud.

Out of the total mass of $5\times10^4\,M_\odot$ in 181 dense
massive clumps in NGC~6334, $3.4\times10^4\,M_\odot$ are contained
within the NGC~6334a region and $1.6\times10^4\,M_\odot$ are outside.
The slope of the CMF of the whole region is dominated by the relatively
low mass clumps in the outer region since they dominate by number.     
The bulk of the cloud mass is located in a few inner clumps which
do not affect significantly the exponent of the derived CMF.   This could 
shed light on the effects of the star formation activity, mass segregation, or
coalescence in the clump mass spectrum within GMCs.
We know that it is more likely to miss low mass clumps due to confusion within the region NGC~6334a than in the outer regions under study.   This bias naturally flattens the slope of the mass spectrum in the inner region.  However, we find in \S 3.6.1 that clumps are not only concentrated by mass towards the center, but also by number, hinting that this result is partly real.    Higher angular resolution observations will be needed to settle this issue.

\subsection{Possible Uncertainties in the CMF}
\label{sec:variate}

In the above analysis we assumed a single temperature for the
whole ensemble of clumps. This is clearly an approximation; the
clumps themselves are not isothermal and the temperature is likely 
to be different from clump to clump.  In addition, we
assumed that all the detected 1.2-mm emission is due to dust
thermal emission. It is possible, however, that some of the 1.2-mm
emission is due to free-free emission from ionized gas. 
In what follows we assess these two assumptions and quantify their 
effects on the derived CMF.

\subsubsection{Temperature}

\begin{figure*}
\plotone{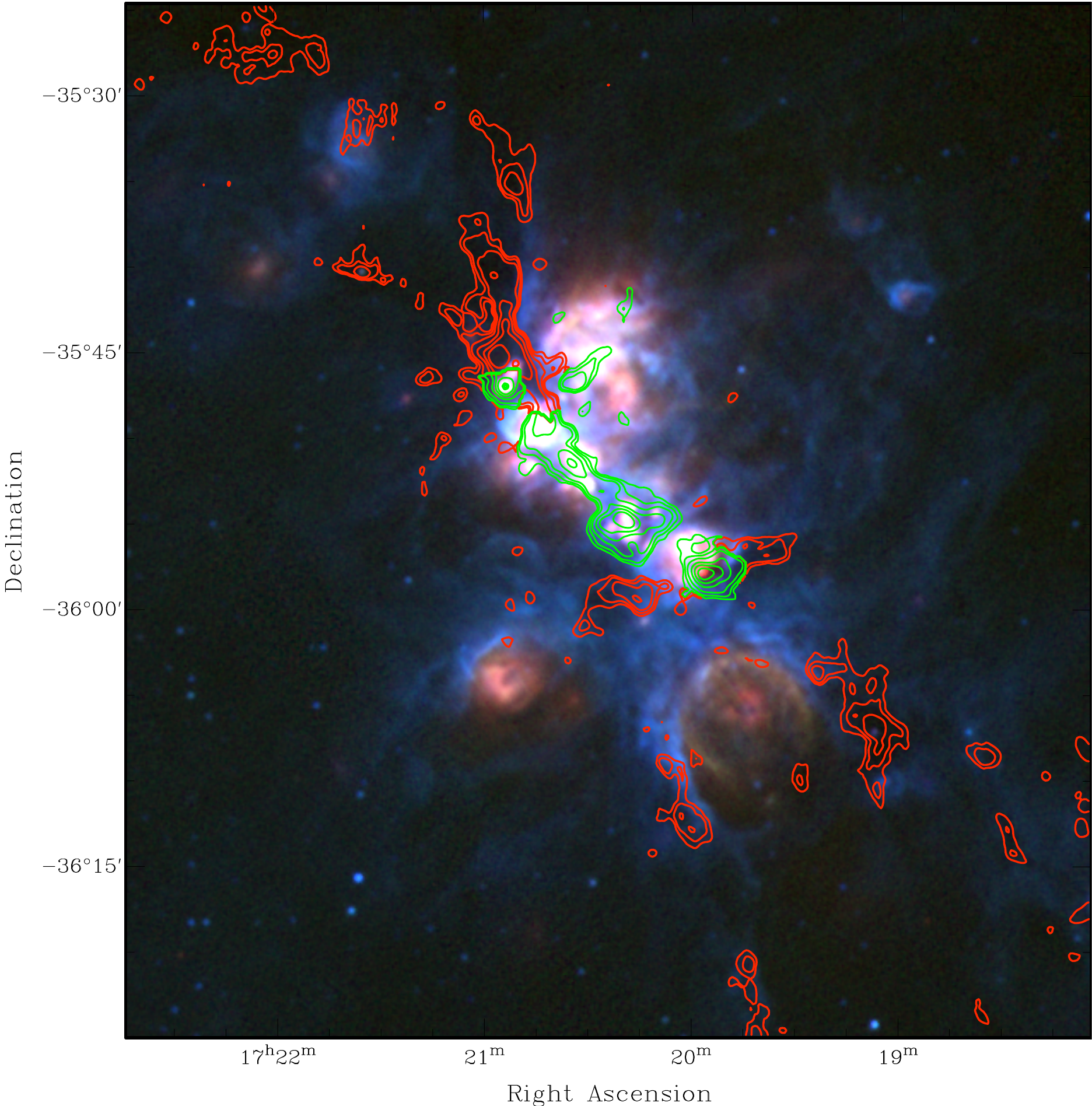}
\caption{Three color image of the MSX emission towards NGC~6334
(blue: Band A, 8.28$\mu$m; green: Band C, 12.13$\mu$m; red: Band E, 21.3$\mu$m).
The countours represent the SIMBA 1.2-mm emission. Green 
contours indicate clumps that are associated with extended or point-like
infrared counterparts. \label{hot_cold_msx}}
\end{figure*}

\begin{figure}
\plotone{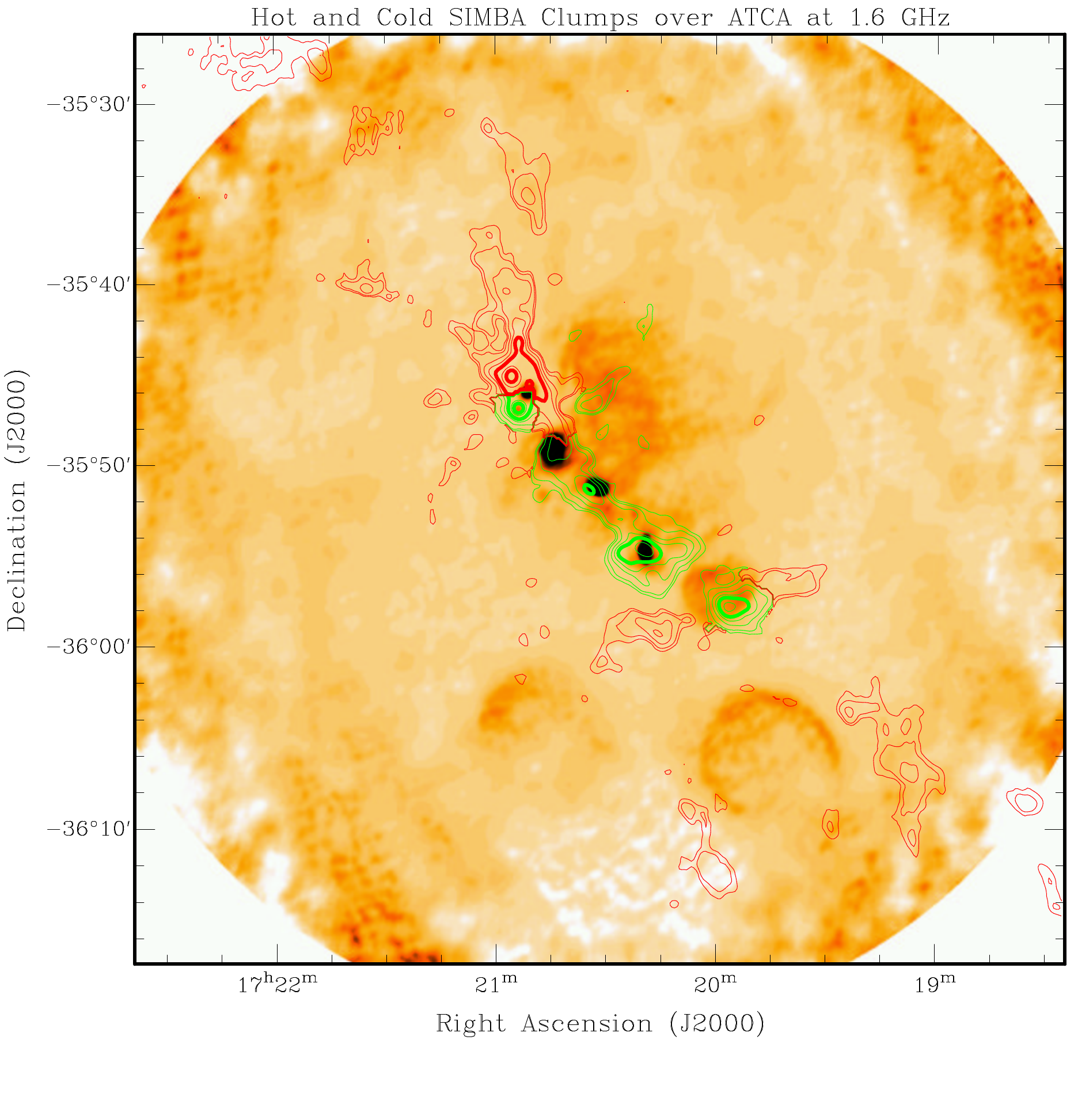} \caption{ATCA radio continuum
emission at 1.6 GHz overlayed with contours of the 1.2-mm emission. 
Again, the red contours indicate cold clumps having no embedded MSX sources
and the green contours indicate warm clumps with embedded MSX sources.
\label{hot_cold_atca}}
\end{figure}

Assuming that all clumps within a GMC are isothermal and have the same 
temperature is clearly a rough approximation. In particular, clumps with 
already formed stars are expected to be warmer than clumps with no signs 
of embedded objects. 
\citet{beu04} have argued that this assumption introduces an uncertainty 
in the derived slope of the mass spectrum. If higher temperatures 
are adopted for the more massive clumps their derived masses will decrease, 
whereas if lower temperatures are adopted for the less massive clumps 
their derived masses will increase: the change would steepen the slope of 
the spectrum.

To study the possible effects of 
temperature differences, we use MSX mid-infrared and/or ATCA cm-continuum 
observations to determine the presence of embedded heating sources
which are likely to be responsible for temperature differences between clumps. 
Figures~\ref{hot_cold_msx} and ~\ref{hot_cold_atca} show, respectively, images 
of the MSX  and ATCA emission overlayed with contours of 
the SIMBA 1.2-mm continuum emission.   We identified by visual inspection 
clumps associated with extended or point-like
infrared counterparts (Figure~\ref{hot_cold_msx}), or clumps associated with 
significant radio continuum (free-free) emission (Figure~\ref{hot_cold_atca}).
These clumps are likely to have embedded sources and are indicated with green contours in both images.    Clumps in red contours appear to be free of embedded infrared or radio sources.

For the purposes of constructing the CMF, the power-law index does not depend 
on the temperature chosen for the whole ensemble of clumps, and the choice of
higher temperatures only displaces the histogram to the left. Any slight variation of  $x$
is due to the intrinsic problem of binning the data. Nevertheless, the choice of two different
temperature can change the shape of the histogram, but essentially depending on the ratio 
between the temperatures chosen rather than the values themselves.

We identified 16 out of 181 cores that appear to be warmer than the 
rest.  For these clumps we adopt a temperature of 34 K, the average 
temperature of massive clumps with embedded IRAS sources as determined 
by \citet{fau04}.  In particular, the temperatures assigned to the cores 
associated with NGC~6334~I and NGC~6334~I(N), of 34~K and 17~K 
respectively, are in accordance with the values given by \cite{gez82}. 
However,~ \citet{san00} argues that source I is much hotter ($T_d\approx100$~K) while
his estimate for I(N) is 30 K, just a factor of two larger than our estimate.
His estimate of the mass of NGC~6334~I of 200 $M_\odot$ Ð in contrast to the $1800\,M_\odot$ in
the present workÐ is not solely explained by the temperature difference but also by the integrated flux. With a resolution of 6''  at 800~$\mu$m,  his estimate for
the size of source I is $10''\times8''$, implying an effective radius $\sim8$ times smaller than our
estimate. Similarly, the definition of \cite{san00} of NGC~6334~I(N) corresponds to the central peak
of the larger clump detected by \citet{gez82}. \citet{san00} obtains a mass of $400\,M_\odot$
for source I(N) after assuming a temperature of 30~K (see \citealp*{san00} and references therein).
Adding up the contributions from the different cores resolved within I(N) and including
the surrounding cloud, \citet{san00} finds that the mass of the whole I(N) region is $\sim2700\,M_\odot$
This value is in agreement with other results found in the literature and is consistent with our result
after taking into account that our choice for the temperature is 17~K. We remark that estimating an exact value of the temperature for each of the clumps of the cloud is not relevant for the statistical analysis we
carry out in the present work.
 
\begin{figure}
\plotone{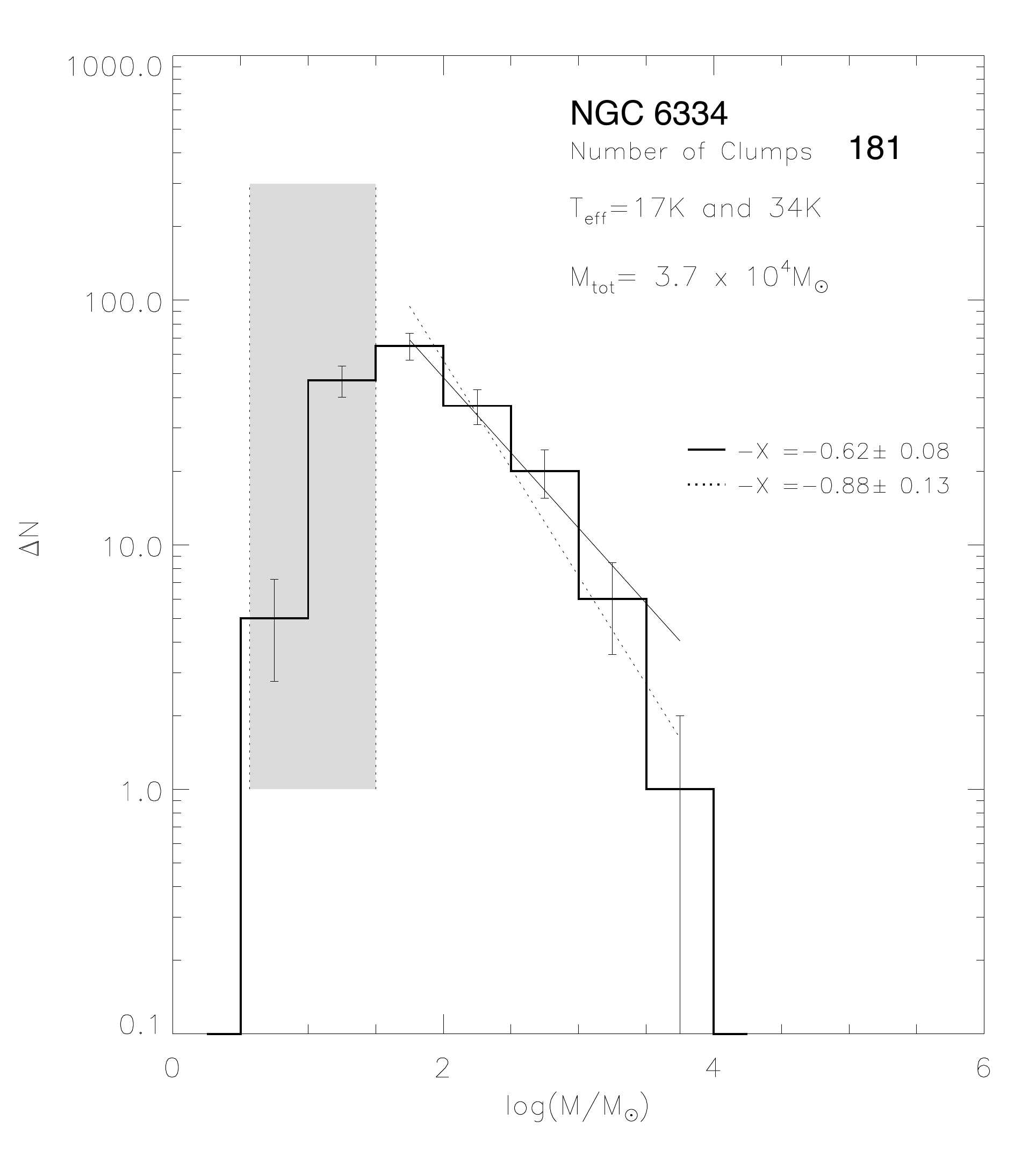}
\caption{Histogram of mass distribution in NGC~6334
for a two-temperature clump ensemble: 17~K and 34~K.
\label{6334histo_2}}
\end{figure}

The newly adopted temperatures imply a repositioning of $\sim$10\% of the 
clumps in the mass histogram. The new mass histogram, made assuming 
a two temperature cloud ensemble, is shown in Figure~\ref{6334histo_2}. 
A linear regression yields a best fit value of $x=0.88\pm 0.13$ (dotted line), but a 
$\chi^2$ fit with Poissonian error bars yields $x=0.62\pm 0.08$ (solid line).
Thus, even though the 
warmer clumps are also preferentially the most massive, they do not concentrate
solely on the most massive bin and do not affect the derived slope of the mass
spectrum significantly.

The repositioning of clumps in the histogram could have a more dramatic effect 
when the number of clumps is considerably smaller. 
For example, the area NGC~6334a
includes 14 of the 16 warmer clumps in NGC~6334 and a total of only 40 clumps
(figure \ref{histograms_abc}).
When using two temperatures in NGC~6334a, the shape of the histogram indeed 
steepens, but the large relative errors yields a slope $x=0.35 \pm 0.21$.   Excluding the last bin, containing only one object,  we obtain a slope of $x=0.09 \pm 0.15$.  
In both cases the slope remains consistent with the value 
$x\sim 0.1$ from a single temperature clump mass distribution.

\subsubsection{Radio Emission Contribution}
Free-free emission is the main continuum contributor at radio
frequencies. At frequencies of hundreds of GHz, the contribution
from \brem  emission is usually neglected and all the detected
emission is assumed to be due to dust. Since massive stars are
being formed within NGC~6334, it is plausible that the 1.2-mm emission
is not completely due to cool dust emission and that the ionized gas
excited by these stars contribute an important amount. \citet{brook05}
studied the Keyhole nebula and found that
the 1.2~mm emission towards the H{\footnotesize{II}}  region
Car-II is strongly
correlated with the 4.8 GHz continuum emission and that there is a lack of
molecular-line emission. They concluded that the 1.2-mm flux from the
components of Car-II arise from free-free emission associated with
ionized gas and not from cool dust emission associated with molecular gas.

Here we make a correction to the observed emission at 250~GHz by free-free 
contamination by estimating the expected ionized gas flux density at 250 GHz 
from the observed flux density at 1.6~GHz.  Assuming that the free-free emission 
is optically thin at both frequencies, the ratio of the emissivities 
is proportional to the ratio of the $e^{-h\nu/kT_e}g_{ff}(\nu,T_e)$ factors
\citep{ryb79}, where $g_{ff}$ is the Gaunt factor. 
The exponential is essentially 1 at both frequencies.
We did not use the usual radio approximation for the
Gaunt factors given by \cite{alten61},  but computed them more precisely 
using quantum mechanical calculations following the work of
Menzel \& Pekeris \citep{men35,som53}.  Table~\ref{tbl-gaunt} 
lists the calculated gaunt factors averaged over a Maxwell-Boltzmann
distribution of velocities with temperatures 
between $7\times10^3$ and $10^4$~K, typical for \htwo regions in massive 
star forming regions \citep{bec00}.   
At 1.6 GHz the computed Gaunt factors are only $\approx 0.3$\% lower than the usual Altenhoff et al.\ approximation.  On the other hand, the calculated Gaunt factors at 250 GHz are typically 15\% smaller than the values obtained from the Altenhoff et al.\ approximation.

\begin{deluxetable}{lllll}
\tabletypesize{\scriptsize} \tablecaption{Gaunt Factors for
Free-Free Emission\label{tbl-gaunt}} \tablewidth{0pt} 
\tablehead{\colhead{Frequency} &  \colhead{T=7000~K} & \colhead{8000~K} & 
	\colhead{9000~K} & \colhead{10000~K} } 
\startdata
1.6 GHz & 5.40327  & 5.5134 & 5.61051 & 5.69735\\
250 GHz &2.79429  & 2.8871 & 2.9702 & 3.04545
\enddata
\end{deluxetable}

Table~\ref{tbl-ff} summarizes the values of  
the estimated free-free emission at 250 GHz from selected clumps
associated with $\htwo$ regions (see Figure~\ref{radio_sources_2}).
We subtracted this flux from the measured 1.2-mm flux density to estimate the 
actual contribution from dust, rederiving the mass of each clump. The 
high mass bins are the most affected, but the best fit exponent ($x=0.88\pm 0.13$) is 
consistent with the previous value.

\begin{deluxetable*}{lrrrrrrrr}
\tabletypesize{\scriptsize} \tablecaption{Radio an MM fluxes [Jy]
for Selected Clumps in Figure~\ref{hot_cold_atca}\label{tbl-ff}}
\tablewidth{0pt} \tablehead{ \colhead{Frequency} & \colhead{cl1} &
\colhead{cl19} & \colhead{cl169} & \colhead{cl11} & \colhead{cl27}
&\colhead{cl7}&\colhead{cl210}&\colhead{cl4} } 
\startdata
ATCA at 1.6 GHz (Free-Free) & 0.59  & 22.77  & 0.45  & 9.42  & 5.28   & 12.61  &3.49  &2.89 \\
Expected Free-Free at 250 GHz & 0.31 &  12.16  &  0.24 &  5.03 & 2.82  & 6.73 & 1.86  &1.54  \\
SIMBA at 250 GHz & 68.38 & 26.78   & 1.13  & 27.12  &  13.69 & 26.72  & 1.34  &59.31  
\enddata
\end{deluxetable*}

\subsubsection{Binning}
\label{sec:binning}

\begin{figure*}
\plotone{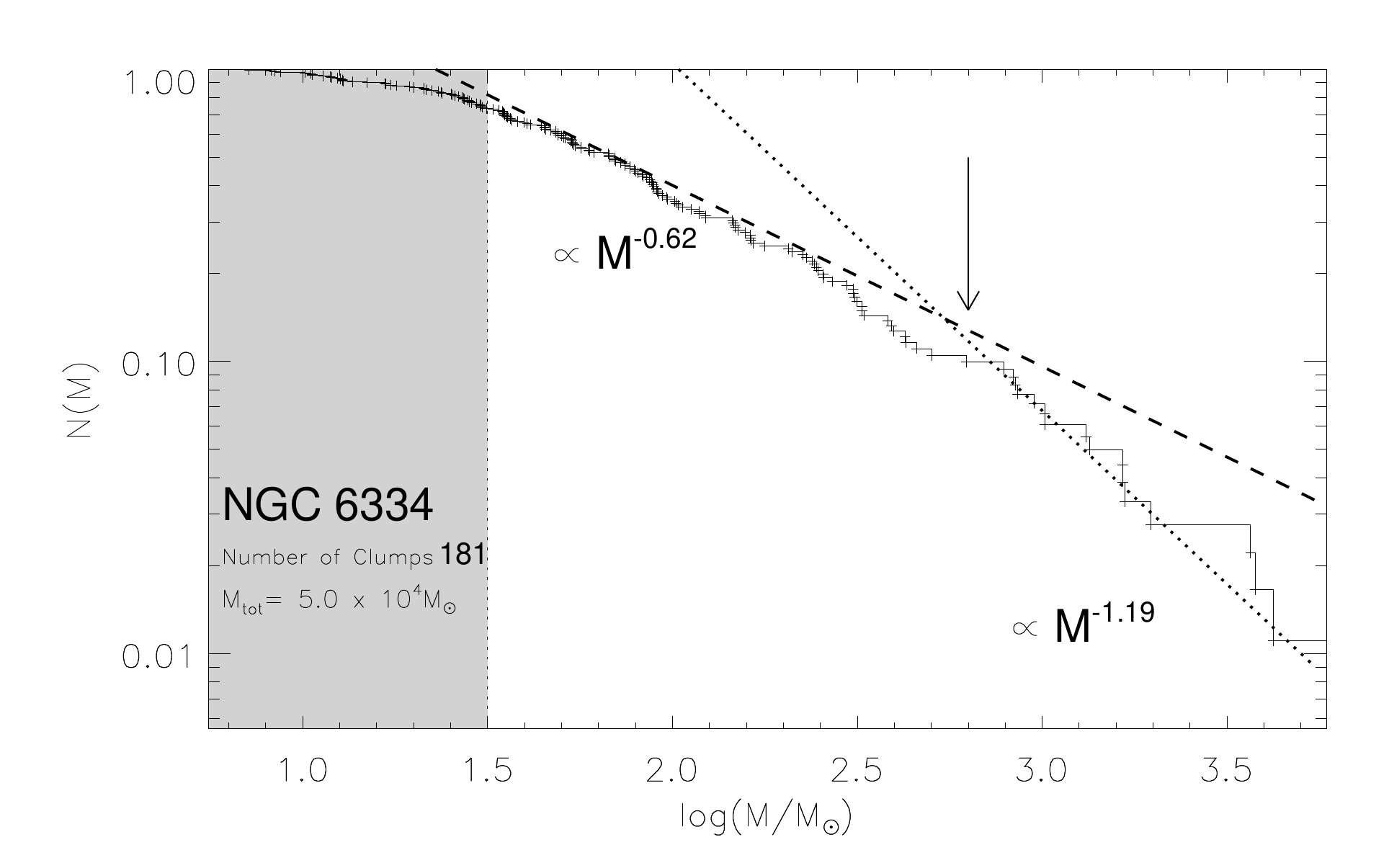}
\caption{
Normalized cumulative mass function ${\cal N}(M)$.  The grey area
corresponds to the region affected by completeness uncertainties
as shown in Figure~\ref{6334histo_1}.   The arrow delimits the top 10\% of the CMF where a power-law fit to the CMF (dotted line) yields an apparent Salpeter-like slope.
\label{cumulative}}
\end{figure*}

Given a distribution of clump masses, the binning process invariably loses
information \citep{roso05}.  The mass histogram can be interpreted as the derivative
of a cumulative number function ${\cal N}(M)$ which counts clumps with mass greater
than $M$,
\begin{equation}
  {\cal N}(M) = \frac{\int_{M}^{\infty} \xi_{\alpha}(m)dm}{\int_{0}^{\infty} \xi_{\alpha}(m)dm} ~.
\end{equation}
Some authors argue in favor of using the cumulative number function ${\cal N}(M)$
\citep{john00b,john01,ker01,tot02} to avoid the loss of information.
If the cumulative number function has a power-law  dependance with mass,
${\cal N}(M) \propto M^{-\gamma}$, then the CMF is also a power law
$\xi_{\alpha}(M) \propto M^{-\gamma-1}$.  From the definitions
in section 3.4, it is clear that $\gamma = x$

However, if there is an upper mass limit in the distribution of cores, then the cumulative mass function is not a power-law , showing considerable curvature at the high mass end.
${\cal N}(M)$ can be approximated by a power-law with index $x=\alpha-1$ only at masses
$M\ll M_{max}$ (or when $M_{max}\rightarrow\infty$).
For finite upper mass limit,
\begin{equation}
{\cal N}(M)=C_1M^{-x}+C_2  ~,
\label{eq-cumul}
\end{equation}
where $C_2$ becomes unimportant for small masses (see \citealp*{roso05,rei06b,li06}).
A power-law form, ${\cal N}(M) \propto M^{-x}$, approximates the true CMF asymptotically towards low masses.  Thus, a power law fit to the CMF must be applied
within a range of masses that avoids {\em both} the incomplete low-mass range
{\em and} the cutoff high-mass range.
When fitting a power-law
function to an observed cumulative mass function in a mass range $M > M_{max}/2$
the slope invariably increases, explaining the
apparent Salpeter-like slopes found in previous studies.   Furthermore, the slope thus obtained is strongly dependent on the break-point chosen to fit the high mass end of the cumulative mass function
and does not reflect the underlying differential clump mass distribution
($\xi_\alpha$ or $\xi_x$).

For sample sizes of $\sim70$ clumps or fewer, binning becomes an important factor
in fitting a power law to differential mass functions and the use of a cumulative mass
function is preferred.   \citet{john00b,john01,john06} and \citet{rei05,rei06a} use
cumulative mass distribution functions to avoid this problem.
Our sample (181 clumps) is large enough to analyse the data using
either the cumulative or the differential mass functions. Figure \ref{cumulative}
plots the normalized cumulative number function of clumps in NGC6334.
A single power-law fit, which we have shown does not represent the true underlying clump mass function, gives ${\cal N}(M) \propto M^{-0.87}$.  The dashed line in Figure \ref{cumulative} shows that the slope $x=0.62$ derived from the histogram (Figure \ref{6334histo_1}) represents a good asymptote to the CMF, as predicted by the theory (Eq.\ \ref{eq-cumul}).
We also show in Figure \ref{cumulative} a fit to the top 10\% of the cloud mass ($\log(M/M_{\odot})>2.8$, indicated by an arrow) which yields a slope $x=1.19$ (dotted line) much closer to Salpeter's value.   The best fit exponent changes to $x=1.37$ for $\log M \ge 2.9$, and to $x=1.47$ for $\log M \ge 3.0$.    Even though these slopes are consistent with Salpeter's IMF, they are an artifact of having an upper clump mass limit in the sample and do not reflect the true clump mass function

In summary, one must be aware of introducing biases when using the high mass range to fit power laws or when fitting broken power laws to the cumulative mass function.
These problems are minimized (but still present) with samples larger than 100 clumps.

\subsection{Spatial Distribution of Clumps}
\label{sec:spatial}

In order to understand the process of fragmentation, we need
to explain how masses are distributed in clumps and how they are
positioned in space. A complete theory of star formation must not
only reproduce the mass function, but it must explain it in all
its physical implications including how clumps, cores and stars
are distributed spatially during the evolution of the GMC (Bonnel et al.\ 2006).
The exhaustive study of
the spatial distribution of young stars in the Taurus region by \citet{gom93}
was extended by other authors and compiled by
\citet{lar95}.  Here we undertake a study the degree of clustering of
clumps in NGC 6334 using a similar approach.

\subsubsection{Clustering and Segregation of Clumps}

The number density of clumps can give us insight about the actual state of
fragmentation and how the clumps are distributed spatially independent of
their mass.   We study the number density of
clumps by means of the Simple Grid\footnote{The Grid
Method consists in binning the two-dimensional space with squares
of side $l$ and then dividing the numbers of sources lying within
each square by the area of it $l^2$, to obtain number density in
units of length$^{-2}$} and the Kernel
Methods\footnote{The Kernel Method \citep{silv86} uses
a kernel function $K$ offering the advantage that the
density distribution is smoothed. In each point $(x,y)$ of
$(\alpha,\delta)$, the kernel density estimator determines the
density due to the contributions of all $n$ data points.}.
Both methods require a free parameter which
determines the ``resolution'' of the number density estimator:
the binning length $D$ in the grid technique, and the smoothing length
$h$ in the Kernel Method, where a kernel $K$ is
defined at each pixel of the map in Figure~\ref{final2_region} by
\begin{equation}
K(x,x_i,y,y_i)=\frac{1}{2\pi}e^{-(x^2+y^2)/2h^2} ~,
\end{equation}
with the kernel density estimator $D$ defined by
\begin{equation}
D(x,y)=\frac{1}{h^2}\sum_{i=1}^nK(x,x_i,y,y_i) ~,
\end{equation}
where $x$ and $y$ are measured in pc ignoring the sky curvature.
The grid bin $l$ was taken to be
2~pc while $h$ was chosen in order to smooth the distribution over
an area $\pi h^2$ similar in size to the area over which the
simple grid technique smoothed the data points ($l^2$).

We applied this analysis to the subregion NGC~6334b which covers and area of
$\sim 330$~pc$^{-2}$.  We find that the probability that the clumps are 
distributed at random within this subregion is  $\sim10^{-5}$, 
which argues in favor of clustered fragmentation at 
scales between 0.1 and 10~pc.  Thus, we conclude that there is spatial segregation 
in clump number in addition to clump mass.

\subsubsection{The Nearest-Neighbor Distribution in NGC~6334}

\begin{figure}
\plotone{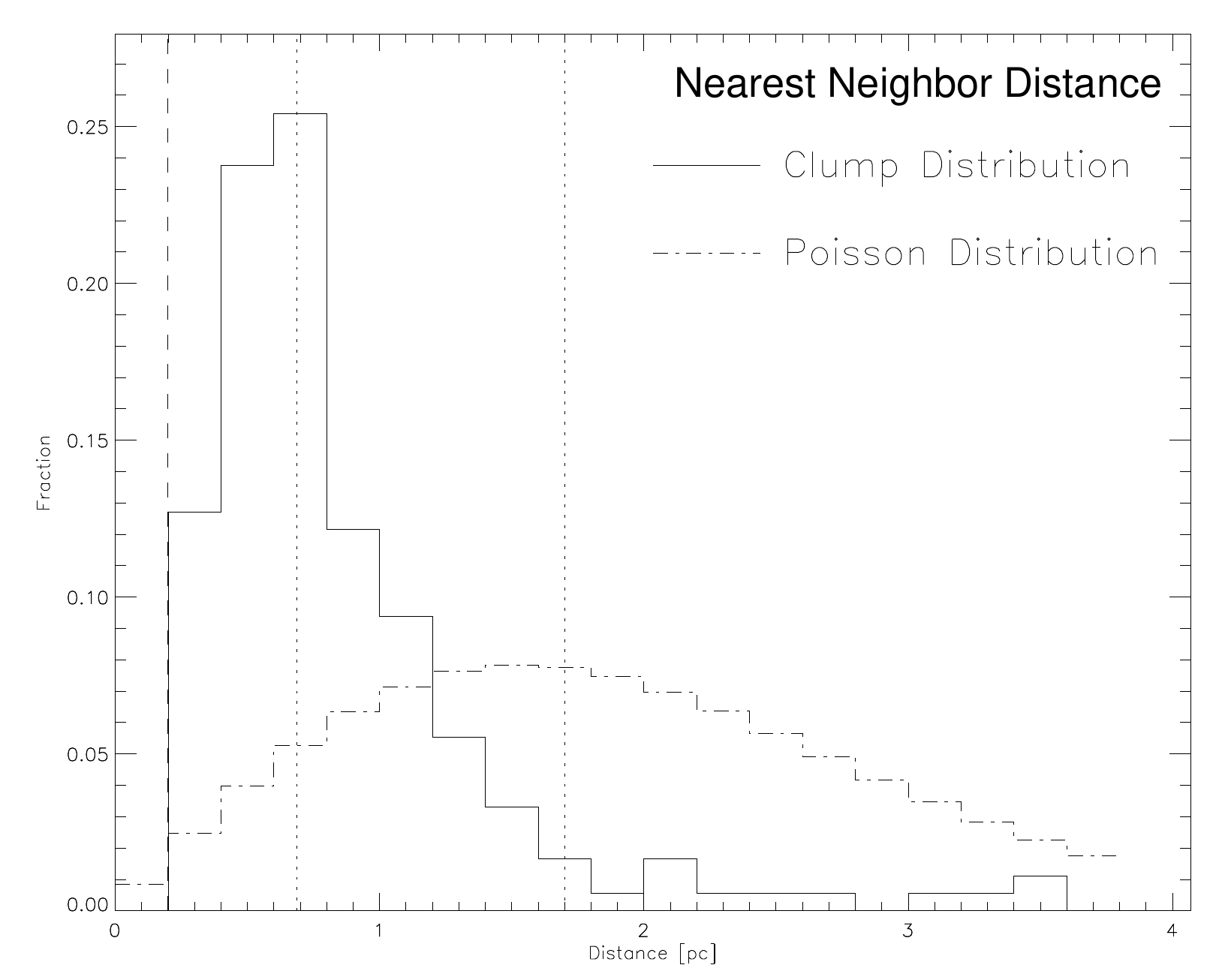}
\caption{
Normalized Nearest Neighbor Distribution after dividing
by 181, the number of neighbor pairs (solid line) and the expected
distribution for random (Poisson) distribution of the same number
of objects as the original sample over an identical area (dot-dashed line). The
medians of the distributions are indicated by dotted lines.
\label{neighbor}}
\end{figure}

We calculate the nearest-neighbor distribution (i.e. the frequency
distribution of the linear distance to the nearest neighbor of
each clump) for the clumps in NGC~6334. 
We neglect those clumps located too close to the edge of
the mapping area but do not exclude them from the total sample
since they can be the nearest neighbor for an inner clump. We
binned the nearest neighbor distances in intervals of 0.2~pc to
construct the histograms shown in Figure~\ref{neighbor}. The nearest
neighbor distribution is strongly skewed to small
separations and very different from the distribution expected
from random positions at the same mean density. The differential
probability of observing at least one event in the interval
$[r_1,r_2]$ which defines a ring surrounding a central source is
\citep{gom93}
\begin{equation}
\Delta P(r_1,r_2,\eta)=e^{-\pi\eta r_1^2}-e^{-\pi\eta r_2^2} ~.
\end{equation}
where $\eta=N/A$ is the average density, in this case given
by $\eta=0.11$~clumps~pc$^{-2}$.
Figure~\ref{neighbor} compares the Poisson
distribution histogram (dashed line) with the normalized
nearest-neighbor distribution for clumps. The median for the
random distribution (1.7~pc) is a factor of $\sim2.5$ greater than
that derived from the actual clump distribution 
(dotted vertical lines in Figure \ref{neighbor}).
We would need to increase the average density $\eta$ by a factor
of 5 to produce a random distribution with the same median
separation as the actual clump distribution. Thus, we conclude
that the median separation of the clump distribution is
significantly smaller than the expected median separation of a
random distribution.

\subsection{The Average Angular Surface Density of Clumps in NGC~6334}

Most of the newly formed stars in nearby regions of star formation are located 
in groups or clusters. The degree of clustering of pre-main sequence stars can 
be obtained by measuring their surface density as a function of angular
distance, $\Sigma(\theta)$, from each star. This surface density can be fit 
by power-laws (Gomez et al.\ 1993); \cite{lar95} found that there is a characteristic 
spatial scale ($\sim 0.04$~pc) where the surface density changes slope. This scale could mark a transition between the regime of cores within molecular clouds 
and protostars within cores. 

We applied the approach followed by \cite{lar95} to the clumps in NGC~6334, 
which have typical sizes $\sim 0.1$~pc and separations of $\sim 1$~pc.
We compute $\Sigma(\theta)$ by taking each clump $k$ and
dividing the surrounding area of the sky into a set of annuli 
of radius $\theta_i$ (with $\theta_i=1.5\theta_{i-1}$), 
and counting the number of companion
clumps $\mathcal{N}_k(\theta_i)$ in each annulus \citep{kit98}.
Then,
\begin{equation}\label{surface2}
\Sigma(\overline{\theta_i})=\frac{1}{N}\frac{\sum_{k=1}^N\mathcal{N}_k(\theta_i)} 
{\pi(\theta_i^2-\theta_{i-1}^2)} ~,  \;\;\;\;\;\;\;\;\;  
\overline{\theta_i}=\frac{(\theta_i+\theta_{i-1})}{2} ~.
\end{equation}

\begin{figure}
\plotone{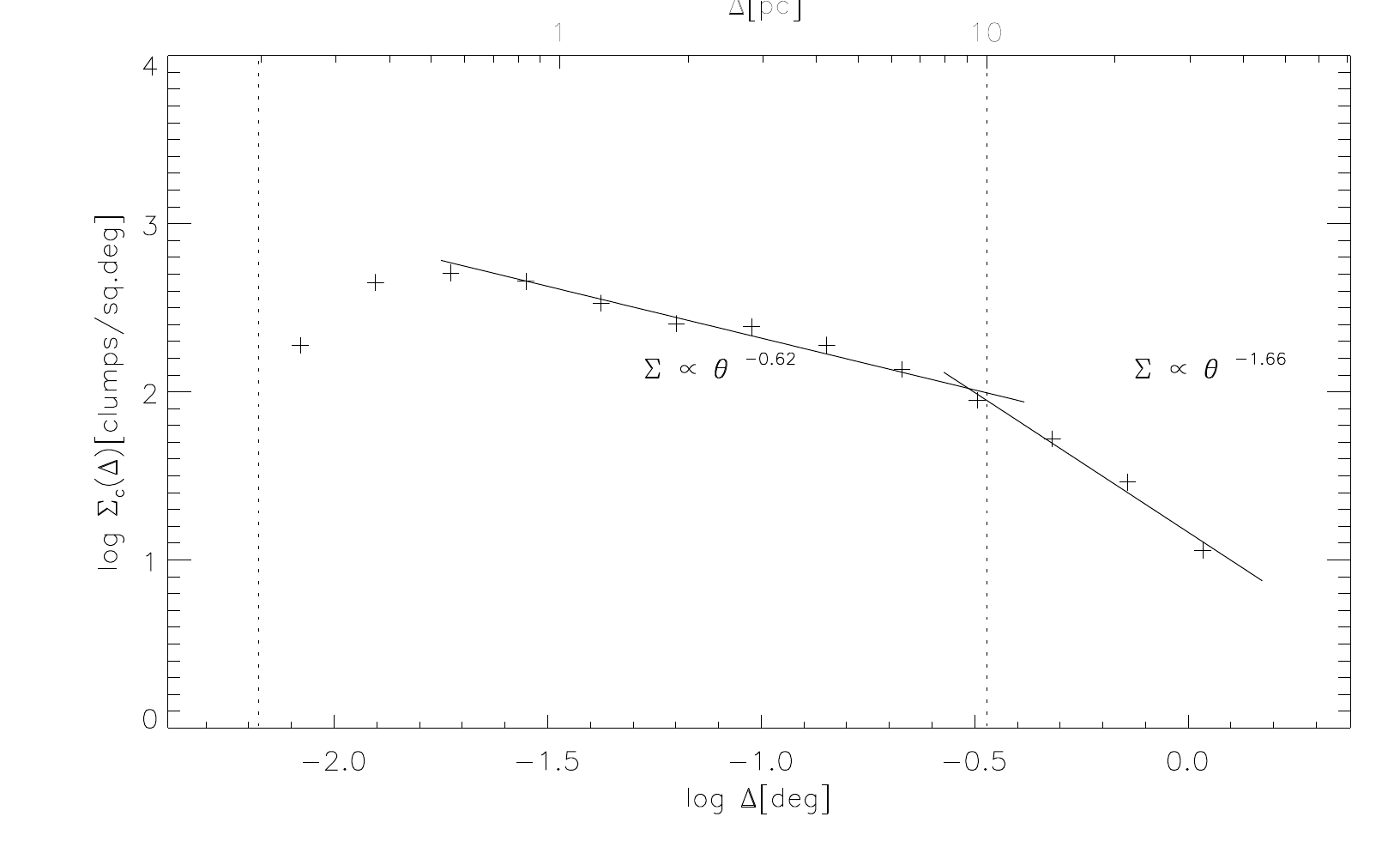}
\caption{
Average Surface
Density as a function of separation in degrees (Equation~\ref{surface2}). 
The vertical dotted lines encompass the limits of validity for the fit: 
the beamsize $\log\theta=-2.7$ ($\sim0.2$~pc), and 
the sampling bias (the width of subregion NGC~6334b) at
$\log\theta=-0.5$ or 10~pc.
\label{surface_density_2}}
\end{figure}

The results are plotted in Figure~\ref{surface_density_2}. 
The surface density is poorly fitted by a single power-law,
but the fit is much improved when using a broken-power law.
However, this reflects a sampling problem at large separations,
since the area observed around NGC~6334 is not square,
Its narrowest part has a width of $\sim 10$~pc, just 
where the power-law apparently breaks. 
The reliable portion of the power-law has a slope of $-0.62$
\begin{eqnarray}
\Sigma_c\propto\theta^{-0.62}&\;\;\;\;\;\;\;&0.6\,\mathrm{pc}\leq
\theta D \leq 10\,\mathrm{pc} ~,
\end{eqnarray}
which is remarkably similar to the one found by
\cite{lar95} for separations larger than 0.04~pc calculated
from young stars in the Taurus-Auriga region:
\begin{eqnarray}
\Sigma_c\propto\theta^{-0.62}&\;\;\;\;\;\;\;&0.04\,\mathrm{pc}\leq
\theta D \leq2.5\,\mathrm{pc} .
\end{eqnarray}
These two results cover two different separation ranges in regions
with widely different physical properties, yet the
surface density distribution is the same. If clustering above the
characteristic length of 0.04~pc maintains a self-similar behavior
up to 10~pc, then the grouping of stars at large separations
(between subclusters for example) could be determined from early
stages of cloud fragmentationÐ before stars are formed.  

\section{DISCUSSION}
\label{sec:discussion}

\paragraph{The CMF and its relation to the IMF.}
The clump mass function is, on one hand tied to the formation and 
evolution of their {\it parent}\ molecular clouds, and on the other to 
the formation of their {\it daughter}\ stars.  Provided that the mechanism of clump 
fragmentation and collapse to form stars is universal, then the IMF should be 
a direct consequence of the CMF.  However, since a single clump 
generally forms multiple stars, we can't expect the distribution of clump masses and stellar masses to have identical functional form.

We find that the mass spectrum of clumps in NGC~6334 has a power-law 
dependence with mass with an index $x = 0.62$ ($dN/d\log M \propto M^{-0.62}$). This 
value is similar to values derived from dust continuum observations for other 
massive star forming regions \citep{ker01,tot02,mook05}.
It is also similar to values derived from molecular line observations,
usually isotopomeric lines of CO, for clouds with similar total mass.
For instance, \citet{noz91} finds $\alpha =1.7$ in Ophiucus North
(mass range of 4--250~$M_\odot$), \citet{stut90} finds the same exponent in
M17SW (mass range of 10--3200~$M_\odot$), and \citet{kram98} reports
$\alpha =1.8$ for NGC~7538 (mass range of 50--3.9$\times10^3$~$M_\odot$).
These results indicate that the clumps in all these massive star forming regions 
are not the direct progenitors of individual stars.
The fact that the massive clumps follow a power-law mass function over a
wide range in mass is remarkable.  The similarity of the
"high mass" CMF slope to the Mass Function of GMC's in our Galaxy
\citep{san85,solo87,pud02} seems to support a hierarchical or
fractal model of the distribution of gas in the Milky Way, where fragmentation
and mass distribution can be interpreted as scale free.

In order to build a complete theory of star formation we must understand 
the process of clustered star formation in clumps satisfying the observed
CMF ($dN/d\log M\propto M^{-0.6}$) and leading to the observed IMF
($dN/d\log M\propto M^{-1.35}$). This has stimulated studies about the 
process of fragmentation of GMCs and clumps both theoretically and observationally.
Hydrodynamic and magnetohydrodynamic numerical calculations suggest that 
turbulence might play a major role in the clump fragmentation process.  
For example \cite{bonn98} suggest that stars are formed through the competitive accretion of 
gas onto proto-cores within molecular clouds. 

Recent sub-millimeter and millimeter continuum observations of low-mass star 
forming regions (e.g.\ Ophiucus, Serpens, Orion B) \citep{mott98,tes98,john00b,john01,beu04} have revealed a CMF with a slope similar to that to the IMF.
The different values obtained from the molecular line observations described 
above and these continuum observations probably reflects the different mass 
ranges sampled. In fact, dust continuum emission observations toward more distant 
massive star forming regions have revealed CMF slopes consistent with a value of -0.6 
(\citealp*{ker01,tot02,mook05}; this work), similar to that obtained from 
molecular line observations.
From observations of a sample of massive clumps toward the massive 
star forming region NGC~7538, \citet{rei05} concluded that a Salpeter-like 
mass function is already established at the earliest stages of star formation. 
This is a surprising result since many of their clumps, with masses of
$10^2-10^3M_\odot$, are still likely to be undergoing the process of fragmentation 
and can not be direct progenitors of individual stars.
Even if the structure in ISM were fractal, self-similarity must break
on small scales, where star formation is taking place.    
We note, however, that the result of \citet{rei05} comes from a fit to the high mass end 
of the cumulative mass function.   We showed in section 3.5.3 that these 
fits are biased towards larger exponents. In addition to molecular line and dust continuum surveys,
extinction maps can be used to map the dense molecular cores in star forming regions. Using this
technique, \citet{alv07} have found a mass spectrum for cores in the Pipe Nubula that is surprisingly similar to the IMF. Their CMF
displaced to higher masses with respect to the stellar IMF only by a factor of 3, suggesting a one-to-one 
mapping from cores to stars with a star formation efficiency of $\sim30$\%. In their case, the cores
detected span a range of masses of 1-10~$M_\odot$ approximately. This clearly indicates a much
smaller scale than the clumps in our work whether determined by the tracer or just the spatial resolution
of the observations. It is  not clear if the scale of the observations determines the observed
fragmentation conditions Ð i.e. the distribution of masses Ð or if the IMF is a result of the complex
evolution of the accreting cores and their interplay with the harboring molecular cloud and other companions. The low star forming activity in the Pipe Nebula has led \citet{al07} to suggest that
this CMF can be determined at early evolutionary stages.    

Assuming that at scales of 0.4~pc and masses from $100M_\odot$
to $5000M_\odot$ the {\sl Blitz} slope ($\equiv 0.6$) is valid, the question that
arises is: What must happen to change the slope from Blitz-like to
Salpeter-like? In the context of gravitational opacity limited
fragmentation, gravitational collapse starts from density
inhomogeneities and proceeds with cooling, which in turn
produces  smaller Jeans masses in the colder regions, favoring 
gravitational fragmentation on small scales. In a strictly self-similar
regime, fragmentation should occur  maintaining the same slope in
the mass spectrum at any scale.  But at some point the gas cores will not be
supported by thermal or non-thermal motions, collapse will occur and 
stars will form, halting the clump fragmentation, while larger and less 
dense clumps will continue fragmenting.  Eventually large clump masses 
will be depleted, and the number of small mass clumps will increase 
steepening the slope to eventually reach a Salpeter value. The exact 
form in which this happens is probably a combination of many of the 
mechanisms proposed.

\paragraph{Will these clumps form stars?}
\label{sec:bound}
The discussion above assumes that all clumps fragment to form stars in order to obtain the IMF from a CMF. However, this might not be the case, in particular for some of the least massive clumps further away from the cloud center. Are these clumps gravitationally bound? Are
they hence likely to collapse or are they only transient structures or
overdensities triggered by turbulent compressive shocks? These questions have yet to be answered,either observationally or theoretically.

Numerical simulations commonly report that many of the lower-mass cores 
formed are not gravitationally bound \citep{kless01,cla05,till05}.   
Furthermore, the mass spectra can be understood as due to purely hydrodynamical
effects without gravity (e.g. \citealp*{cla06}). As 
\citet{pado02} remark, the mass spectra resulting
from turbulent fragmentation is different from the one that
considers collapsing or unstable cores.  Only the latter form
stars. These calculations show that many of the clumps could be
transient structures; indeed a significant fraction of the cores
end up re-expanding rather than collapsing \citep{vaz05,nak05}.
This implies that fragmentation is not sufficient to trigger star
formation.  If supersonic turbulence generates the initial density
enhancements from which cores develop, then these cores might not
necessarily approach hydrostatic equilibrium at any point in their
evolution \citep{ball99}. 

Can we distinguish observationally between bound and transient clumps? To
estimate if clumps are likely or not to collapse, we need to know more
than its mass.  A clump's column density is
important as a diagnostic of whether the clump is likely to
collapse \citep{pud02}.  Values higher than
$N\simeq10^{22}$~cm$^{-2}$ are observed for cores with embedded
sources.    The virial parameter, $M_\mathrm{vir}/M$,  where $M$ is a mass derived from column density and $M_\mathrm{vir}$ is the virial mass derived from the cloud radius and velocity dispersion\citep{bert92}, is known to have values close to 1.0 in star forming clouds \citep{oni96,yone05}.   However, given the observational uncertainties it is not yet clear if this parameter is a good diagnostic for star forming vs. transient clumps. 

\paragraph{Preferred spatial scale.}
\label{sec:hierarchy}
As \cite{lar95} found in Taurus, there might exist a
preferred scale of star formation at which clustering changes.
This length can be related to the Jeans length and the Jeans mass. The surface number density of the clumps as a function of separation can reveal
a characteristic spatial scale, marking a transition between clumps and cores. At separations larger than 0.6~pc and smaller than 10~pc, we find
the same slope for the power-law fit to the surface density of
companions as the protostars in Taurus do at large separations:
$\Sigma\propto\theta^{-0.62}$. This suggests that large
separations in stellar systems are determined by the position of
their progenitor clumps. 

\section{CONCLUSIONS}
We made observations of the 1.2-mm dust emission toward
the Giant Molecular Cloud NGC~6334 using the SIMBA bolometer at the SEST.  The main results 
and conclusions presented in this paper are summarized as follows. 

We find 181 clumps, distributed in an area of $\sim 2.0$ square degrees
centered on the main filament, which harbors most of the
star-forming activity. The clumps range in size from 0.1 to 0.9~pc, with 
a median of 0.35~pc. This range is similar to that found by \citet{fau04} 
and \citet{plum97} for clumps in different high mass star forming regions.
The clump masses, assuming they are isothermal, range from 3~M$_\odot$ to
$6\times10^3\,M_\odot$, with a  completeness limit of
 $\sim 30$~M$_\odot$ (assuming $T_{d}=17$~K).

The Clump Mass Function (CMF) is well fit with a power-law 
dependence with mass with an index $x=0.62\pm0.07$ ($dN/d\log M \propto M^{-x}$) 
in the mass range between $30$~M$_\odot$ to $6\times 10^3$~M$_\odot$.  
The slope differs from the stellar IMF slope, indicating that clumps are not 
direct progenitors of stars.  Therefore other processesÐ besides
fragmentationÐ must be important in setting up the IMF from the CMF.

We assessed possible effects on the derived slope of the CMF due to 
changes on the temperature assumptions and due to the contribution of 
free-free emission from ionized gas to the 1.2-mm emission.
Although $\lesssim10\%$ of the clumps are likely to be
significantly warmer than 17~K and are associated with 
regions of ionized gas, the correction for temperature 
and free-free emission has little effects on the derived slope.

We investigated possible differences in the value of $x$ among 
different sub-regions of NGC~6334. We find that the slope is significantly shallower 
toward the central filament ($x\approx 0.1$), which contains the most 
massive clumps and represents the minimum of the gravitational potential
in the GMC.  As we cover more extended regions, with clumps not actively forming stars,
the slope steepens ($\approx 0.5,\,0.8$), revealing that the
bulk of the clumps are located in the outer areas of the molecular
cloud and that these low-mass clumps predominantly 
determine the shape of the mass function.

We caution about the power-law fitting procedures to the mass
function.  The differential CMF 
is sensitive to bin size and to low-number statistics 
in the last bin (high-mass end) as well as to completeness
limits in the low-mass end.  On the other hand, fitting a power-law
to the high mass end of the cumulative CMF is incorrect 
due to its high mass cutoff.   Both the low-mass and high-mass ends of the cumulative CMF must be
avoided in fitting power-laws.

The spatial analysis performed on the two-dimensional
distribution of clumps reveals that they are not distributed
randomly. They are concentrated toward the center of the
filament, indicating not only a segregation in mass but also a
segregation in number which could suggest a possible coalescence
of massive clumps towards the gravitational potential minimum. In
addition, we study the surface density of companions as a
function of separation.   This is well fit by a power-law with a similar 
exponent to
the one found for proto-stars in Taurus at large angular separations.  
This suggests that the position of stars in clusters is
determined in the fragmentation and star formation stage rather
than after dynamical relaxation.

\acknowledgments

We thank Sim\'on Casassus and Dieter Nurnberger for helpful comments.   
D.J.M., D.M., G.G., and D.R. gratefully acknowledge support from the Chilean 
{\sl Centro de Astrof\'\i sica} FONDAP No.\ 15010003.

\clearpage
\LongTables
\begin{deluxetable*}{llrcrrllrl}
\tablewidth{0pt} \tabletypesize{\scriptsize} \tablecaption{Clump
Properties in NGC 6334}\label{clump_properties} \tablehead{
\colhead{Clump} & \colhead{$\alpha$} & \colhead{$\delta$} &
\colhead{$I_{peak}$} & \colhead{$S_\nu$}&
\colhead{$R_\mathrm{eff}$}  & \colhead{$M_\mathrm{1.2mm}$\tablenotemark{a}} 
& \colhead{$M$} & \colhead{$n$
\tablenotemark{b}} & \colhead{region}\\
 \colhead{} & \colhead{(J2000)}      & \colhead{J2000}        & \colhead{[Jy
beam$^{-1}$]} & \colhead{[Jy] }     & \colhead{[pc]} &
\colhead{[$M_\odot$]} &\colhead{[$M_\odot$]} & \colhead{[cm$^{-3}$]} &
\colhead{}}
 \startdata

cl~{\bf1} & 17$^h$20$^m$53.9$^s$ & -35$^\mathrm{\circ}$47' 1.3'' & 17.47 & 
 68.38 & 0.59 &  4.22$\times10^{3}$&  1.74$\times10^{3}$&  3.48$\times10^{4}$
  & {\bf NGC 6334}{\it(6334a)}{\it(6334b)} \\
cl~2 & 17$^h$20$^m$55.9$^s$ & -35$^\mathrm{\circ}$45'17.3'' & 14.92 &  95.17 & 
0.65 &  5.88$\times10^{3}$&  5.88$\times10^{3}$&  8.86$\times10^{4}$  & 
{\bf NGC 6334}{\it(6334a)}{\it(6334b)} \\
cl~{\bf3} & 17$^h$20$^m$19.8$^s$ & -35$^\mathrm{\circ}$54'45.7'' &  8.92 & 
 61.08 & 0.62 &  3.77$\times10^{3}$&  1.56$\times10^{3}$&  2.70$\times10^{4}$
  & {\bf NGC 6334}{\it(6334a)}{\it(6334b)} \\
cl~{\bf4} & 17$^h$19$^m$56.7$^s$ & -35$^\mathrm{\circ}$57'58.0'' &  7.15 & 
 59.31 & 0.86 &  3.66$\times10^{3}$&  1.47$\times10^{3}$&  9.69$\times10^{3}$
  & {\bf NGC 6334}{\it(6334a)}{\it(6334b)} \\
cl~{\bf6} & 17$^h$17$^m$ 1.9$^s$ & -36$^\mathrm{\circ}$20'53.9'' &  4.13 & 
 31.86 & 0.92 &  1.97$\times10^{3}$&  8.12$\times10^{2}$&  4.37$\times10^{3}$
  & {\bf NGC 6334}{\it(6334c)} \\
cl~{\bf7} & 17$^h$20$^m$24.4$^s$ & -35$^\mathrm{\circ}$55' 1.9'' &  3.58 & 
 26.72 & 0.59 &  1.65$\times10^{3}$&  5.10$\times10^{2}$&  1.02$\times10^{4}$
  & {\bf NGC 6334}{\it(6334a)}{\it(6334b)} \\
cl~8 & 17$^h$20$^m$49.9$^s$ & -35$^\mathrm{\circ}$45' 9.4'' &  2.91 &  16.45 & 
0.36 &  1.02$\times10^{3}$&  1.02$\times10^{3}$&  9.44$\times10^{4}$  & 
{\bf NGC 6334}{\it(6334a)}{\it(6334b)} \\
cl~9 & 17$^h$20$^m$48.0$^s$ & -35$^\mathrm{\circ}$45'57.2'' &  2.74 &  21.73 & 
0.45 &  1.34$\times10^{3}$&  1.34$\times10^{3}$&  6.38$\times10^{4}$  & 
{\bf NGC 6334}{\it(6334a)}{\it(6334b)} \\
cl~10 & 17$^h$20$^m$53.9$^s$ & -35$^\mathrm{\circ}$43'25.3'' &  2.62 &  13.90 & 
0.45 &  8.59$\times10^{2}$&  8.59$\times10^{2}$&  4.08$\times10^{4}$  & 
{\bf NGC 6334}{\it(6334a)}{\it(6334b)} \\
cl~{\bf11} & 17$^h$20$^m$34.2$^s$ & -35$^\mathrm{\circ}$51'33.8'' &  2.30 & 
 27.12 & 0.71 &  1.68$\times10^{3}$&  5.63$\times10^{2}$&  6.54$\times10^{3}$
  & {\bf NGC 6334}{\it(6334a)}{\it(6334b)} \\
cl~{\bf14} & 17$^h$23$^m$16.8$^s$ & -34$^\mathrm{\circ}$48'43.6'' &  1.98 & 
 13.54 & 0.68 &  8.36$\times10^{2}$&  3.45$\times10^{2}$&  4.55$\times10^{3}$
  & {\bf NGC 6334}{\it(6334c)} \\
cl~{\bf15} & 17$^h$20$^m$10.5$^s$ & -35$^\mathrm{\circ}$54'54.0'' &  1.70 & 
 21.30 & 0.74 &  1.32$\times10^{3}$&  5.43$\times10^{2}$&  5.58$\times10^{3}$
  & {\bf NGC 6334}{\it(6334a)}{\it(6334b)} \\
cl~16 & 17$^h$20$^m$43.4$^s$ & -35$^\mathrm{\circ}$47'49.6'' &  1.68 &  16.44 & 
0.50 &  1.02$\times10^{3}$&  1.02$\times10^{3}$&  3.32$\times10^{4}$  & 
{\bf NGC 6334}{\it(6334a)}{\it(6334b)} \\
cl~{\bf19} & 17$^h$20$^m$42.7$^s$ & -35$^\mathrm{\circ}$49'17.4'' &  1.36 & 
 26.78 & 0.68 &  1.65$\times10^{3}$&  3.73$\times10^{2}$&  4.92$\times10^{3}$
  & {\bf NGC 6334}{\it(6334a)}{\it(6334b)} \\
cl~20 & 17$^h$20$^m$53.9$^s$ & -35$^\mathrm{\circ}$42'21.2'' &  1.29 &   6.93 & 
0.36 &  4.28$\times10^{2}$&  4.28$\times10^{2}$&  3.97$\times10^{4}$  & 
{\bf NGC 6334}{\it(6334a)}{\it(6334b)} \\
cl~23 & 17$^h$20$^m$15.8$^s$ & -35$^\mathrm{\circ}$59'25.8'' &  1.10 &  15.41 & 
0.74 &  9.52$\times10^{2}$&  9.52$\times10^{2}$&  9.77$\times10^{3}$  & 
{\bf NGC 6334}{\it(6334a)}{\it(6334b)} \\
cl~25 & 17$^h$20$^m$57.1$^s$ & -35$^\mathrm{\circ}$40'29.3'' &  1.08 &   6.89 & 
0.39 &  4.25$\times10^{2}$&  4.25$\times10^{2}$&  3.11$\times10^{4}$  & 
{\bf NGC 6334}{\it(6334a)}{\it(6334b)} \\
cl~{\bf27} & 17$^h$20$^m$25.7$^s$ & -35$^\mathrm{\circ}$53'10.0'' &  1.02 & 
 13.69 & 0.59 &  8.45$\times10^{2}$&  2.77$\times10^{2}$&  5.56$\times10^{3}$
  & {\bf NGC 6334}{\it(6334a)}{\it(6334b)} \\
cl~29 & 17$^h$20$^m$55.8$^s$ & -35$^\mathrm{\circ}$41'41.3'' &  0.94 &   5.27 & 
0.36 &  3.25$\times10^{2}$&  3.25$\times10^{2}$&  3.02$\times10^{4}$  & 
{\bf NGC 6334}{\it(6334a)}{\it(6334b)} \\
cl~{\bf30} & 17$^h$20$^m$32.9$^s$ & -35$^\mathrm{\circ}$46'45.8'' &  0.93 & 
  7.41 & 0.56 &  4.58$\times10^{2}$&  1.89$\times10^{2}$&  4.42$\times10^{3}$
  & {\bf NGC 6334}{\it(6334a)}{\it(6334b)} \\
cl~33 & 17$^h$20$^m$51.9$^s$ & -35$^\mathrm{\circ}$40'29.3'' &  0.81 &   5.27 & 
0.39 &  3.26$\times10^{2}$&  3.26$\times10^{2}$&  2.38$\times10^{4}$  & 
{\bf NGC 6334}{\it(6334a)}{\it(6334b)} \\
cl~34 & 17$^h$21$^m$35.2$^s$ & -35$^\mathrm{\circ}$40'19.6'' &  0.81 &   5.04 & 
0.53 &  3.11$\times10^{2}$&  3.11$\times10^{2}$&  8.57$\times10^{3}$  & 
{\bf NGC 6334}{\it(6334b)} \\
cl~35 & 17$^h$19$^m$ 7.8$^s$ & -36$^\mathrm{\circ}$ 7' 1.2'' &  0.77 &  12.77 & 
0.77 &  7.89$\times10^{2}$&  7.89$\times10^{2}$&  7.20$\times10^{3}$  & 
{\bf NGC 6334}{\it(6334b)} \\
cl~37 & 17$^h$20$^m$50.5$^s$ & -35$^\mathrm{\circ}$35'25.4'' &  0.74 &  10.11 & 
0.74 &  6.24$\times10^{2}$&  6.24$\times10^{2}$&  6.41$\times10^{3}$  & 
{\bf NGC 6334}{\it(6334b)} \\
cl~42 & 17$^h$20$^m$56.4$^s$ & -35$^\mathrm{\circ}$39'41.0'' &  0.68 &   5.00 & 
0.39 &  3.09$\times10^{2}$&  3.09$\times10^{2}$&  2.25$\times10^{4}$  & 
{\bf NGC 6334}{\it(6334a)}{\it(6334b)} \\
cl~43 & 17$^h$17$^m$21.5$^s$ & -36$^\mathrm{\circ}$ 8'31.6'' &  0.68 &   4.16 & 
0.56 &  2.57$\times10^{2}$&  2.57$\times10^{2}$&  6.01$\times10^{3}$  & 
{\bf NGC 6334}{\it(6334c)} \\
cl~45 & 17$^h$17$^m$49.2$^s$ & -36$^\mathrm{\circ}$ 9'13.7'' &  0.64 &   6.43 & 
0.53 &  3.97$\times10^{2}$&  3.97$\times10^{2}$&  1.09$\times10^{4}$  & 
{\bf NGC 6334}{\it(6334c)} \\
cl~48 & 17$^h$19$^m$23.7$^s$ & -36$^\mathrm{\circ}$ 3'41.8'' &  0.63 &   5.11 & 
0.56 &  3.16$\times10^{2}$&  3.16$\times10^{2}$&  7.38$\times10^{3}$  & 
{\bf NGC 6334}{\it(6334b)} \\
cl~51 & 17$^h$17$^m$43.2$^s$ & -36$^\mathrm{\circ}$10'33.2'' &  0.57 &   5.34 & 
0.53 &  3.30$\times10^{2}$&  3.30$\times10^{2}$&  9.08$\times10^{3}$  & 
{\bf NGC 6334}{\it(6334c)} \\
cl~52 & 17$^h$22$^m$34.1$^s$ & -35$^\mathrm{\circ}$13'11.6'' &  0.57 &   8.13 & 
0.74 &  5.02$\times10^{2}$&  5.02$\times10^{2}$&  5.16$\times10^{3}$  & 
{\bf NGC 6334}{\it(6334c)} \\
cl~54 & 17$^h$22$^m$28.1$^s$ & -35$^\mathrm{\circ}$ 8'56.0'' &  0.55 &   6.33 & 
0.68 &  3.91$\times10^{2}$&  3.91$\times10^{2}$&  5.16$\times10^{3}$  & 
{\bf NGC 6334}{\it(6334c)} \\
cl~56 & 17$^h$17$^m$ 7.2$^s$ & -36$^\mathrm{\circ}$19' 2.3'' &  0.52 &   6.19 & 
0.65 &  3.82$\times10^{2}$&  3.82$\times10^{2}$&  5.76$\times10^{3}$  & 
{\bf NGC 6334}{\it(6334c)} \\
cl~57 & 17$^h$22$^m$ 0.5$^s$ & -35$^\mathrm{\circ}$27'38.2'' &  0.50 &   2.68 & 
0.36 &  1.65$\times10^{2}$&  1.65$\times10^{2}$&  1.54$\times10^{4}$  & 
{\bf NGC 6334}{\it(6334b)} \\
cl~58 & 17$^h$23$^m$25.9$^s$ & -34$^\mathrm{\circ}$48'18.7'' &  0.49 &   3.64 & 
0.45 &  2.25$\times10^{2}$&  2.25$\times10^{2}$&  1.07$\times10^{4}$  & 
{\bf NGC 6334}{\it(6334c)} \\
cl~66 & 17$^h$22$^m$51.6$^s$ & -34$^\mathrm{\circ}$52'54.1'' &  0.45 &   1.57 & 
0.36 &  9.69$\times10^{1}$&  9.69$\times10^{1}$&  9.00$\times10^{3}$  & 
{\bf NGC 6334}{\it(6334c)} \\
cl~74 & 17$^h$20$^m$ 0.0$^s$ & -36$^\mathrm{\circ}$12'54.0'' &  0.42 &   4.14 & 
0.50 &  2.56$\times10^{2}$&  2.56$\times10^{2}$&  8.36$\times10^{3}$  & 
{\bf NGC 6334}{\it(6334b)} \\
cl~75 & 17$^h$19$^m$44.9$^s$ & -35$^\mathrm{\circ}$56'21.8'' &  0.42 &   3.97 & 
0.47 &  2.45$\times10^{2}$&  2.45$\times10^{2}$&  9.60$\times10^{3}$  & 
{\bf NGC 6334}{\it(6334a)}{\it(6334b)} \\
cl~76 & 17$^h$20$^m$40.1$^s$ & -35$^\mathrm{\circ}$45'57.6'' &  0.42 &   2.54 & 
0.42 &  1.57$\times10^{2}$&  1.57$\times10^{2}$&  9.19$\times10^{3}$  & 
{\bf NGC 6334}{\it(6334a)}{\it(6334b)} \\
cl~77 & 17$^h$20$^m$ 2.6$^s$ & -36$^\mathrm{\circ}$12' 6.1'' &  0.42 &   3.92 & 
0.53 &  2.42$\times10^{2}$&  2.42$\times10^{2}$&  6.66$\times10^{3}$  & 
{\bf NGC 6334}{\it(6334b)} \\
cl~79 & 17$^h$19$^m$12.5$^s$ & -36$^\mathrm{\circ}$ 6'13.3'' &  0.41 &   4.39 & 
0.53 &  2.71$\times10^{2}$&  2.71$\times10^{2}$&  7.46$\times10^{3}$  & 
{\bf NGC 6334}{\it(6334b)} \\
cl~81 & 17$^h$19$^m$38.3$^s$ & -35$^\mathrm{\circ}$56'53.9'' &  0.41 &   4.99 & 
0.56 &  3.08$\times10^{2}$&  3.08$\times10^{2}$&  7.21$\times10^{3}$  & 
{\bf NGC 6334}{\it(6334b)} \\
cl~84 & 17$^h$23$^m$29.2$^s$ & -34$^\mathrm{\circ}$48'34.2'' &  0.40 &   1.91 & 
0.36 &  1.18$\times10^{2}$&  1.18$\times10^{2}$&  1.10$\times10^{4}$  & 
{\bf NGC 6334}{\it(6334c)} \\
cl~86 & 17$^h$21$^m$ 1.7$^s$ & -35$^\mathrm{\circ}$39' 1.1'' &  0.39 &   3.86 & 
0.45 &  2.38$\times10^{2}$&  2.38$\times10^{2}$&  1.13$\times10^{4}$  & 
{\bf NGC 6334}{\it(6334a)}{\it(6334b)} \\
cl~89 & 17$^h$18$^m$34.8$^s$ & -36$^\mathrm{\circ}$ 8'44.2'' &  0.37 &   3.33 & 
0.53 &  2.06$\times10^{2}$&  2.06$\times10^{2}$&  5.66$\times10^{3}$  & 
{\bf NGC 6334}{\it(6334c)} \\
cl~92 & 17$^h$20$^m$ 7.3$^s$ & -36$^\mathrm{\circ}$ 9'10.1'' &  0.35 &   1.46 & 
0.33 &  9.04$\times10^{1}$&  9.04$\times10^{1}$&  1.09$\times10^{4}$  & 
{\bf NGC 6334}{\it(6334b)} \\
cl~95 & 17$^h$19$^m$43.4$^s$ & -36$^\mathrm{\circ}$20'46.0'' &  0.34 &   2.39 & 
0.47 &  1.47$\times10^{2}$&  1.47$\times10^{2}$&  5.78$\times10^{3}$  & 
{\bf NGC 6334}{\it(6334c)} \\
cl~96 & 17$^h$21$^m$ 7.7$^s$ & -35$^\mathrm{\circ}$43' 8.8'' &  0.33 &   1.91 & 
0.39 &  1.18$\times10^{2}$&  1.18$\times10^{2}$&  8.61$\times10^{3}$  & 
{\bf NGC 6334}{\it(6334a)}{\it(6334b)} \\
cl~104 & 17$^h$17$^m$59.0$^s$ & -36$^\mathrm{\circ}$13'22.1'' &  0.31 &   1.25
 & 0.33 &  7.69$\times10^{1}$&  7.69$\times10^{1}$&  9.27$\times10^{3}$  & 
{\bf NGC 6334}{\it(6334c)} \\
cl~106 & 17$^h$16$^m$33.8$^s$ & -36$^\mathrm{\circ}$27'31.0'' &  0.31 &   1.64
 & 0.42 &  1.02$\times10^{2}$&  1.02$\times10^{2}$&  5.94$\times10^{3}$  & 
{\bf NGC 6334}{\it(6334c)} \\
cl~108 & 17$^h$21$^m$ 2.4$^s$ & -35$^\mathrm{\circ}$42'29.2'' &  0.30 &   2.63
 & 0.45 &  1.62$\times10^{2}$&  1.62$\times10^{2}$&  7.72$\times10^{3}$  & 
{\bf NGC 6334}{\it(6334a)}{\it(6334b)} \\
cl~109 & 17$^h$22$^m$ 5.1$^s$ & -35$^\mathrm{\circ}$27'37.8'' &  0.30 &   1.41
 & 0.33 &  8.68$\times10^{1}$&  8.68$\times10^{1}$&  1.05$\times10^{4}$  & 
{\bf NGC 6334}{\it(6334b)} \\
cl~110 & 17$^h$21$^m$49.3$^s$ & -35$^\mathrm{\circ}$27'23.0'' &  0.30 &   3.73
 & 0.56 &  2.30$\times10^{2}$&  2.30$\times10^{2}$&  5.39$\times10^{3}$  & 
{\bf NGC 6334}{\it(6334b)} \\
cl~111 & 17$^h$21$^m$ 4.4$^s$ & -35$^\mathrm{\circ}$46'21.0'' &  0.29 &   1.44
 & 0.36 &  8.92$\times10^{1}$&  8.92$\times10^{1}$&  8.28$\times10^{3}$  & 
{\bf NGC 6334}{\it(6334a)}{\it(6334b)} \\
cl~112 & 17$^h$22$^m$30.7$^s$ & -35$^\mathrm{\circ}$11'12.1'' &  0.29 &   3.42
 & 0.56 &  2.11$\times10^{2}$&  2.11$\times10^{2}$&  4.94$\times10^{3}$  & 
{\bf NGC 6334}{\it(6334c)} \\
cl~113 & 17$^h$21$^m$16.2$^s$ & -35$^\mathrm{\circ}$46'28.6'' &  0.28 &   1.47
 & 0.36 &  9.06$\times10^{1}$&  9.06$\times10^{1}$&  8.41$\times10^{3}$  & 
{\bf NGC 6334}{\it(6334b)} \\
cl~114 & 17$^h$19$^m$40.8$^s$ & -36$^\mathrm{\circ}$24'54.0'' &  0.27 &   2.64
 & 0.50 &  1.63$\times10^{2}$&  1.63$\times10^{2}$&  5.32$\times10^{3}$  & 
{\bf NGC 6334}{\it(6334c)} \\
cl~115 & 17$^h$22$^m$ 7.7$^s$ & -35$^\mathrm{\circ}$27'45.7'' &  0.27 &   2.40
 & 0.50 &  1.48$\times10^{2}$&  1.48$\times10^{2}$&  4.84$\times10^{3}$  & 
{\bf NGC 6334}{\it(6334b)} \\
cl~117 & 17$^h$21$^m$ 2.3$^s$ & -35$^\mathrm{\circ}$31'32.9'' &  0.27 &   0.97
 & 0.30 &  5.97$\times10^{1}$&  5.97$\times10^{1}$&  9.57$\times10^{3}$  & 
{\bf NGC 6334}{\it(6334b)} \\
cl~118 & 17$^h$22$^m$33.6$^s$ & -34$^\mathrm{\circ}$58'23.5'' &  0.26 &   4.01
 & 0.65 &  2.48$\times10^{2}$&  2.48$\times10^{2}$&  3.74$\times10^{3}$  & 
{\bf NGC 6334}{\it(6334c)} \\
cl~120 & 17$^h$19$^m$ 6.5$^s$ & -36$^\mathrm{\circ}$10'37.2'' &  0.26 &   2.88
 & 0.53 &  1.78$\times10^{2}$&  1.78$\times10^{2}$&  4.89$\times10^{3}$  & 
{\bf NGC 6334}{\it(6334b)} \\
cl~122 & 17$^h$22$^m$52.5$^s$ & -35$^\mathrm{\circ}$16'38.3'' &  0.26 &   1.48
 & 0.39 &  9.11$\times10^{1}$&  9.11$\times10^{1}$&  6.65$\times10^{3}$  & 
{\bf NGC 6334}{\it(6334c)} \\
cl~123 & 17$^h$20$^m$31.0$^s$ & -36$^\mathrm{\circ}$ 1' 1.9'' &  0.26 &   2.43
 & 0.50 &  1.50$\times10^{2}$&  1.50$\times10^{2}$&  4.90$\times10^{3}$  & 
{\bf NGC 6334}{\it(6334b)} \\
cl~127 & 17$^h$21$^m$36.3$^s$ & -35$^\mathrm{\circ}$32' 3.5'' &  0.25 &   2.35
 & 0.47 &  1.45$\times10^{2}$&  1.45$\times10^{2}$&  5.70$\times10^{3}$  & 
{\bf NGC 6334}{\it(6334b)} \\
cl~128 & 17$^h$19$^m$ 7.9$^s$ & -36$^\mathrm{\circ}$ 4'29.3'' &  0.25 &   4.79
 & 0.65 &  2.96$\times10^{2}$&  2.96$\times10^{2}$&  4.46$\times10^{3}$  & 
{\bf NGC 6334}{\it(6334b)} \\
cl~129 & 17$^h$21$^m$28.6$^s$ & -35$^\mathrm{\circ}$40'52.0'' &  0.24 &   0.59
 & 0.24 &  3.64$\times10^{1}$&  3.64$\times10^{1}$&  1.14$\times10^{4}$  & 
{\bf NGC 6334}{\it(6334b)} \\
cl~131 & 17$^h$17$^m$58.5$^s$ & -36$^\mathrm{\circ}$ 9' 6.1'' &  0.24 &   2.00
 & 0.42 &  1.23$\times10^{2}$&  1.23$\times10^{2}$&  7.21$\times10^{3}$  & 
{\bf NGC 6334}{\it(6334c)} \\
cl~132 & 17$^h$18$^m$27.2$^s$ & -36$^\mathrm{\circ}$25'39.7'' &  0.24 &   1.28
 & 0.36 &  7.90$\times10^{1}$&  7.90$\times10^{1}$&  7.34$\times10^{3}$  & 
{\bf NGC 6334}{\it(6334c)} \\
cl~133 & 17$^h$20$^m$ 7.9$^s$ & -35$^\mathrm{\circ}$58'45.8'' &  0.23 &   0.92
 & 0.30 &  5.66$\times10^{1}$&  5.66$\times10^{1}$&  9.09$\times10^{3}$  & 
{\bf NGC 6334}{\it(6334a)}{\it(6334b)} \\
cl~134 & 17$^h$22$^m$ 3.7$^s$ & -35$^\mathrm{\circ}$24'42.1'' &  0.23 &   1.99
 & 0.45 &  1.23$\times10^{2}$&  1.23$\times10^{2}$&  5.85$\times10^{3}$  & 
{\bf NGC 6334}{\it(6334b)} \\
cl~135 & 17$^h$22$^m$ 1.8$^s$ & -35$^\mathrm{\circ}$26'18.2'' &  0.23 &   1.09
 & 0.33 &  6.71$\times10^{1}$&  6.71$\times10^{1}$&  8.09$\times10^{3}$  & 
{\bf NGC 6334}{\it(6334b)} \\
cl~136 & 17$^h$21$^m$32.4$^s$ & -35$^\mathrm{\circ}$31'31.8'' &  0.23 &   1.49
 & 0.42 &  9.19$\times10^{1}$&  9.19$\times10^{1}$&  5.37$\times10^{3}$  & 
{\bf NGC 6334}{\it(6334b)} \\
cl~138 & 17$^h$19$^m$28.9$^s$ & -36$^\mathrm{\circ}$ 9'57.6'' &  0.23 &   1.56
 & 0.42 &  9.64$\times10^{1}$&  9.64$\times10^{1}$&  5.64$\times10^{3}$  & 
{\bf NGC 6334}{\it(6334b)} \\
cl~140 & 17$^h$21$^m$33.1$^s$ & -35$^\mathrm{\circ}$32' 3.8'' &  0.22 &   0.74
 & 0.27 &  4.55$\times10^{1}$&  4.55$\times10^{1}$&  1.00$\times10^{4}$  & 
{\bf NGC 6334}{\it(6334b)} \\
cl~141 & 17$^h$18$^m$26.8$^s$ & -36$^\mathrm{\circ}$14'11.8'' &  0.22 &   1.24
 & 0.36 &  7.68$\times10^{1}$&  7.68$\times10^{1}$&  7.14$\times10^{3}$  & 
{\bf NGC 6334}{\it(6334c)} \\
cl~142 & 17$^h$21$^m$ 6.4$^s$ & -35$^\mathrm{\circ}$47'33.0'' &  0.22 &   0.56
 & 0.24 &  3.47$\times10^{1}$&  3.47$\times10^{1}$&  1.09$\times10^{4}$  & 
{\bf NGC 6334}{\it(6334a)}{\it(6334b)} \\
cl~143 & 17$^h$17$^m$58.4$^s$ & -36$^\mathrm{\circ}$10' 2.3'' &  0.22 &   1.30
 & 0.36 &  8.06$\times10^{1}$&  8.06$\times10^{1}$&  7.49$\times10^{3}$  & 
{\bf NGC 6334}{\it(6334c)} \\
cl~145 & 17$^h$21$^m$11.7$^s$ & -35$^\mathrm{\circ}$50' 4.6'' &  0.22 &   0.86
 & 0.30 &  5.34$\times10^{1}$&  5.34$\times10^{1}$&  8.57$\times10^{3}$  & 
{\bf NGC 6334}{\it(6334b)} \\
cl~146 & 17$^h$20$^m$ 3.3$^s$ & -36$^\mathrm{\circ}$ 9'25.9'' &  0.22 &   1.21
 & 0.33 &  7.45$\times10^{1}$&  7.45$\times10^{1}$&  8.98$\times10^{3}$  & 
{\bf NGC 6334}{\it(6334b)} \\
cl~147 & 17$^h$17$^m$10.9$^s$ & -36$^\mathrm{\circ}$26'30.8'' &  0.22 &   1.72
 & 0.47 &  1.06$\times10^{2}$&  1.06$\times10^{2}$&  4.17$\times10^{3}$  & 
{\bf NGC 6334}{\it(6334c)} \\
cl~148 & 17$^h$21$^m$ 1.7$^s$ & -35$^\mathrm{\circ}$37'25.0'' &  0.22 &   1.82
 & 0.42 &  1.12$\times10^{2}$&  1.12$\times10^{2}$&  6.57$\times10^{3}$  & 
{\bf NGC 6334}{\it(6334b)} \\
cl~149 & 17$^h$21$^m$29.1$^s$ & -35$^\mathrm{\circ}$31'31.8'' &  0.22 &   0.67
 & 0.27 &  4.13$\times10^{1}$&  4.13$\times10^{1}$&  9.08$\times10^{3}$  & 
{\bf NGC 6334}{\it(6334b)} \\
cl~151 & 17$^h$23$^m$13.0$^s$ & -34$^\mathrm{\circ}$49'40.1'' &  0.22 &   0.99
 & 0.30 &  6.13$\times10^{1}$&  6.13$\times10^{1}$&  9.83$\times10^{3}$  & 
{\bf NGC 6334}{\it(6334c)} \\
cl~152 & 17$^h$17$^m$41.3$^s$ & -36$^\mathrm{\circ}$ 9'13.0'' &  0.21 &   1.13
 & 0.33 &  6.97$\times10^{1}$&  6.97$\times10^{1}$&  8.41$\times10^{3}$  & 
{\bf NGC 6334}{\it(6334c)} \\
cl~{\bf153} & 17$^h$20$^m$19.7$^s$ & -35$^\mathrm{\circ}$42'29.9'' &  0.21 & 
  1.17 & 0.39 &  7.24$\times10^{1}$&  2.99$\times10^{1}$&  2.18$\times10^{3}$
  & {\bf NGC 6334}{\it(6334b)} \\
cl~155 & 17$^h$18$^m$28.8$^s$ & -36$^\mathrm{\circ}$12'51.8'' &  0.21 &   1.29
 & 0.36 &  7.97$\times10^{1}$&  7.97$\times10^{1}$&  7.40$\times10^{3}$  & 
{\bf NGC 6334}{\it(6334c)} \\
cl~156 & 17$^h$18$^m$15.9$^s$ & -36$^\mathrm{\circ}$26'51.4'' &  0.20 &   1.14
 & 0.39 &  7.01$\times10^{1}$&  7.01$\times10^{1}$&  5.12$\times10^{3}$  & 
{\bf NGC 6334}{\it(6334c)} \\
cl~{\bf159} & 17$^h$20$^m$31.6$^s$ & -35$^\mathrm{\circ}$48'29.9'' &  0.20 & 
  0.46 & 0.24 &  2.82$\times10^{1}$&  1.17$\times10^{1}$&  3.65$\times10^{3}$
  & {\bf NGC 6334}{\it(6334a)}{\it(6334b)} \\
cl~161 & 17$^h$19$^m$58.7$^s$ & -36$^\mathrm{\circ}$ 8'46.0'' &  0.20 &   0.56
 & 0.24 &  3.48$\times10^{1}$&  3.48$\times10^{1}$&  1.09$\times10^{4}$  & 
{\bf NGC 6334}{\it(6334b)} \\
cl~163 & 17$^h$22$^m$14.2$^s$ & -35$^\mathrm{\circ}$26'49.2'' &  0.20 &   2.37
 & 0.53 &  1.46$\times10^{2}$&  1.46$\times10^{2}$&  4.02$\times10^{3}$  & 
{\bf NGC 6334}{\it(6334b)} \\
cl~164 & 17$^h$16$^m$13.2$^s$ & -36$^\mathrm{\circ}$28'25.0'' &  0.20 &   0.57
 & 0.24 &  3.51$\times10^{1}$&  3.51$\times10^{1}$&  1.10$\times10^{4}$  & 
{\bf NGC 6334}{\it(6334c)} \\
cl~165 & 17$^h$18$^m$ 7.5$^s$ & -36$^\mathrm{\circ}$17'54.6'' &  0.20 &   1.21
 & 0.36 &  7.44$\times10^{1}$&  7.44$\times10^{1}$&  6.91$\times10^{3}$  & 
{\bf NGC 6334}{\it(6334c)} \\
cl~167 & 17$^h$18$^m$30.5$^s$ & -36$^\mathrm{\circ}$25'23.9'' &  0.20 &   0.82
 & 0.30 &  5.08$\times10^{1}$&  5.08$\times10^{1}$&  8.16$\times10^{3}$  & 
{\bf NGC 6334}{\it(6334c)} \\
cl~{\bf169} & 17$^h$20$^m$48.0$^s$ & -35$^\mathrm{\circ}$51'49.3'' &  0.19 & 
  1.13 & 0.36 &  6.99$\times10^{1}$&  2.26$\times10^{1}$&  2.10$\times10^{3}$
  & {\bf NGC 6334}{\it(6334a)}{\it(6334b)} \\
cl~171 & 17$^h$22$^m$36.2$^s$ & -34$^\mathrm{\circ}$57'19.4'' &  0.19 &   0.73
 & 0.30 &  4.51$\times10^{1}$&  4.51$\times10^{1}$&  7.24$\times10^{3}$  & 
{\bf NGC 6334}{\it(6334c)} \\
cl~175 & 17$^h$23$^m$19.3$^s$ & -35$^\mathrm{\circ}$16'59.5'' &  0.19 &   0.86
 & 0.33 &  5.32$\times10^{1}$&  5.32$\times10^{1}$&  6.42$\times10^{3}$  & 
{\bf NGC 6334}{\it(6334c)} \\
cl~176 & 17$^h$16$^m$40.4$^s$ & -36$^\mathrm{\circ}$27'39.6'' &  0.19 &   1.66
 & 0.45 &  1.02$\times10^{2}$&  1.02$\times10^{2}$&  4.86$\times10^{3}$  & 
{\bf NGC 6334}{\it(6334c)} \\
cl~179 & 17$^h$20$^m$55.2$^s$ & -35$^\mathrm{\circ}$48'45.4'' &  0.18 &   0.35
 & 0.21 &  2.16$\times10^{1}$&  2.16$\times10^{1}$&  1.01$\times10^{4}$  & 
{\bf NGC 6334}{\it(6334a)}{\it(6334b)} \\
cl~180 & 17$^h$17$^m$37.5$^s$ & -36$^\mathrm{\circ}$26'24.7'' &  0.18 &   1.38
 & 0.42 &  8.53$\times10^{1}$&  8.53$\times10^{1}$&  4.99$\times10^{3}$  & 
{\bf NGC 6334}{\it(6334c)} \\
cl~181 & 17$^h$16$^m$53.0$^s$ & -36$^\mathrm{\circ}$27'49.0'' &  0.18 &   1.52
 & 0.45 &  9.39$\times10^{1}$&  9.39$\times10^{1}$&  4.47$\times10^{3}$  & 
{\bf NGC 6334}{\it(6334c)} \\
cl~182 & 17$^h$21$^m$56.5$^s$ & -35$^\mathrm{\circ}$26'42.4'' &  0.18 &   1.35
 & 0.36 &  8.33$\times10^{1}$&  8.33$\times10^{1}$&  7.73$\times10^{3}$  & 
{\bf NGC 6334}{\it(6334b)} \\
cl~184 & 17$^h$21$^m$ 5.8$^s$ & -35$^\mathrm{\circ}$48'52.9'' &  0.18 &   0.64
 & 0.27 &  3.97$\times10^{1}$&  3.97$\times10^{1}$&  8.73$\times10^{3}$  & 
{\bf NGC 6334}{\it(6334b)} \\
cl~185 & 17$^h$20$^m$44.0$^s$ & -35$^\mathrm{\circ}$39'49.3'' &  0.18 &   0.48
 & 0.24 &  2.95$\times10^{1}$&  2.95$\times10^{1}$&  9.23$\times10^{3}$  & 
{\bf NGC 6334}{\it(6334b)} \\
cl~186 & 17$^h$18$^m$ 7.1$^s$ & -36$^\mathrm{\circ}$ 7'54.8'' &  0.18 &   1.68
 & 0.45 &  1.04$\times10^{2}$&  1.04$\times10^{2}$&  4.94$\times10^{3}$  & 
{\bf NGC 6334}{\it(6334c)} \\
cl~188 & 17$^h$22$^m$ 8.5$^s$ & -35$^\mathrm{\circ}$10'57.7'' &  0.18 &   0.49
 & 0.24 &  3.06$\times10^{1}$&  3.06$\times10^{1}$&  9.58$\times10^{3}$  & 
{\bf NGC 6334}{\it(6334c)} \\
cl~189 & 17$^h$22$^m$30.5$^s$ & -35$^\mathrm{\circ}$25'19.9'' &  0.18 &   0.46
 & 0.24 &  2.87$\times10^{1}$&  2.87$\times10^{1}$&  8.98$\times10^{3}$  & 
{\bf NGC 6334}{\it(6334b)} \\
cl~190 & 17$^h$18$^m$ 7.0$^s$ & -36$^\mathrm{\circ}$12'50.8'' &  0.18 &   0.78
 & 0.30 &  4.83$\times10^{1}$&  4.83$\times10^{1}$&  7.75$\times10^{3}$  & 
{\bf NGC 6334}{\it(6334c)} \\
cl~192 & 17$^h$23$^m$ 6.1$^s$ & -35$^\mathrm{\circ}$13'32.9'' &  0.17 &   0.81
 & 0.33 &  5.01$\times10^{1}$&  5.01$\times10^{1}$&  6.04$\times10^{3}$  & 
{\bf NGC 6334}{\it(6334c)} \\
cl~194 & 17$^h$20$^m$56.4$^s$ & -35$^\mathrm{\circ}$32'21.1'' &  0.17 &   2.63
 & 0.53 &  1.62$\times10^{2}$&  1.62$\times10^{2}$&  4.47$\times10^{3}$  & 
{\bf NGC 6334}{\it(6334b)} \\
cl~197 & 17$^h$17$^m$58.4$^s$ & -36$^\mathrm{\circ}$12'18.4'' &  0.17 &   0.58
 & 0.27 &  3.56$\times10^{1}$&  3.56$\times10^{1}$&  7.84$\times10^{3}$  & 
{\bf NGC 6334}{\it(6334c)} \\
cl~198 & 17$^h$20$^m$54.0$^s$ & -35$^\mathrm{\circ}$50'37.3'' &  0.17 &   0.59
 & 0.27 &  3.66$\times10^{1}$&  3.66$\times10^{1}$&  8.06$\times10^{3}$  & 
{\bf NGC 6334}{\it(6334a)}{\it(6334b)} \\
cl~200 & 17$^h$19$^m$42.1$^s$ & -36$^\mathrm{\circ}$22'37.9'' &  0.17 &   1.09
 & 0.36 &  6.74$\times10^{1}$&  6.74$\times10^{1}$&  6.26$\times10^{3}$  & 
{\bf NGC 6334}{\it(6334c)} \\
cl~201 & 17$^h$19$^m$48.2$^s$ & -35$^\mathrm{\circ}$47'42.0'' &  0.17 &   0.44
 & 0.24 &  2.72$\times10^{1}$&  2.72$\times10^{1}$&  8.52$\times10^{3}$  & 
{\bf NGC 6334}{\it(6334b)} \\
cl~203 & 17$^h$20$^m$ 2.6$^s$ & -36$^\mathrm{\circ}$ 0' 6.1'' &  0.17 &   0.96
 & 0.36 &  5.93$\times10^{1}$&  5.93$\times10^{1}$&  5.51$\times10^{3}$  & 
{\bf NGC 6334}{\it(6334a)}{\it(6334b)} \\
cl~204 & 17$^h$21$^m$18.8$^s$ & -35$^\mathrm{\circ}$42'12.2'' &  0.16 &   0.58
 & 0.27 &  3.58$\times10^{1}$&  3.58$\times10^{1}$&  7.87$\times10^{3}$  & 
{\bf NGC 6334}{\it(6334a)}{\it(6334b)} \\
cl~205 & 17$^h$18$^m$ 5.7$^s$ & -36$^\mathrm{\circ}$11'30.5'' &  0.16 &   1.43
 & 0.45 &  8.82$\times10^{1}$&  8.82$\times10^{1}$&  4.19$\times10^{3}$  & 
{\bf NGC 6334}{\it(6334c)} \\
cl~208 & 17$^h$19$^m$38.1$^s$ & -36$^\mathrm{\circ}$23'33.7'' &  0.16 &   0.57
 & 0.27 &  3.51$\times10^{1}$&  3.51$\times10^{1}$&  7.74$\times10^{3}$  & 
{\bf NGC 6334}{\it(6334c)} \\
cl~209 & 17$^h$18$^m$35.9$^s$ & -36$^\mathrm{\circ}$23'48.1'' &  0.16 &   0.88
 & 0.36 &  5.45$\times10^{1}$&  5.45$\times10^{1}$&  5.06$\times10^{3}$  & 
{\bf NGC 6334}{\it(6334c)} \\
cl~210 & 17$^h$19$^m$58.0$^s$ & -35$^\mathrm{\circ}$55'58.1'' &  0.16 &   1.34
 & 0.42 &  8.28$\times10^{1}$& *****$ $& *****$ $  & 
{\bf NGC 6334}{\it(6334a)}{\it(6334b)} \\
cl~213 & 17$^h$22$^m$ 2.4$^s$ & -35$^\mathrm{\circ}$23'30.1'' &  0.16 &   1.09
 & 0.39 &  6.74$\times10^{1}$&  6.74$\times10^{1}$&  4.92$\times10^{3}$  & 
{\bf NGC 6334}{\it(6334b)} \\
cl~214 & 17$^h$20$^m$11.2$^s$ & -36$^\mathrm{\circ}$14'13.9'' &  0.15 &   0.20
 & 0.15 &  1.24$\times10^{1}$&  1.24$\times10^{1}$&  1.59$\times10^{4}$  & 
{\bf NGC 6334}{\it(6334c)} \\
cl~{\bf215} & 17$^h$20$^m$38.8$^s$ & -35$^\mathrm{\circ}$43' 1.6'' &  0.15 & 
  0.55 & 0.27 &  3.38$\times10^{1}$&  1.40$\times10^{1}$&  3.08$\times10^{3}$
  & {\bf NGC 6334}{\it(6334a)}{\it(6334b)} \\
cl~216 & 17$^h$17$^m$33.2$^s$ & -36$^\mathrm{\circ}$13'36.5'' &  0.15 &   0.53
 & 0.27 &  3.27$\times10^{1}$&  3.27$\times10^{1}$&  7.19$\times10^{3}$  & 
{\bf NGC 6334}{\it(6334c)} \\
cl~217 & 17$^h$18$^m$ 0.9$^s$ & -36$^\mathrm{\circ}$17'22.2'' &  0.15 &   0.87
 & 0.33 &  5.37$\times10^{1}$&  5.37$\times10^{1}$&  6.48$\times10^{3}$  & 
{\bf NGC 6334}{\it(6334c)} \\
cl~218 & 17$^h$21$^m$56.6$^s$ & -35$^\mathrm{\circ}$28'18.5'' &  0.15 &   0.92
 & 0.33 &  5.66$\times10^{1}$&  5.66$\times10^{1}$&  6.83$\times10^{3}$  & 
{\bf NGC 6334}{\it(6334b)} \\
cl~219 & 17$^h$23$^m$52.1$^s$ & -34$^\mathrm{\circ}$51'51.5'' &  0.15 &   0.49
 & 0.27 &  3.01$\times10^{1}$&  3.01$\times10^{1}$&  6.63$\times10^{3}$  & 
{\bf NGC 6334}{\it(6334c)} \\
cl~220 & 17$^h$21$^m$17.0$^s$ & -35$^\mathrm{\circ}$53' 8.5'' &  0.15 &   0.27
 & 0.18 &  1.68$\times10^{1}$&  1.68$\times10^{1}$&  1.25$\times10^{4}$  & 
{\bf NGC 6334}{\it(6334b)} \\
cl~223 & 17$^h$21$^m$23.4$^s$ & -35$^\mathrm{\circ}$41' 0.2'' &  0.15 &   0.16
 & 0.15 &  1.01$\times10^{1}$&  1.01$\times10^{1}$&  1.30$\times10^{4}$  & 
{\bf NGC 6334}{\it(6334a)}{\it(6334b)} \\
cl~226 & 17$^h$22$^m$38.5$^s$ & -35$^\mathrm{\circ}$29'19.3'' &  0.15 &   0.44
 & 0.24 &  2.70$\times10^{1}$&  2.70$\times10^{1}$&  8.46$\times10^{3}$  & 
{\bf NGC 6334}{\it(6334b)} \\
cl~227 & 17$^h$20$^m$46.8$^s$ & -35$^\mathrm{\circ}$59'25.4'' &  0.15 &   0.57
 & 0.27 &  3.50$\times10^{1}$&  3.50$\times10^{1}$&  7.70$\times10^{3}$  & 
{\bf NGC 6334}{\it(6334b)} \\
cl~229 & 17$^h$20$^m$52.7$^s$ & -35$^\mathrm{\circ}$59'49.2'' &  0.14 &   0.49
 & 0.27 &  3.03$\times10^{1}$&  3.03$\times10^{1}$&  6.68$\times10^{3}$  & 
{\bf NGC 6334}{\it(6334b)} \\
cl~231 & 17$^h$23$^m$11.4$^s$ & -35$^\mathrm{\circ}$14'52.4'' &  0.14 &   0.83
 & 0.33 &  5.12$\times10^{1}$&  5.12$\times10^{1}$&  6.17$\times10^{3}$  & 
{\bf NGC 6334}{\it(6334c)} \\
cl~232 & 17$^h$21$^m$47.7$^s$ & -35$^\mathrm{\circ}$39'47.2'' &  0.14 &   0.32
 & 0.21 &  1.95$\times10^{1}$&  1.95$\times10^{1}$&  9.10$\times10^{3}$  & 
{\bf NGC 6334}{\it(6334b)} \\
cl~233 & 17$^h$23$^m$13.3$^s$ & -35$^\mathrm{\circ}$15'32.0'' &  0.14 &   0.76
 & 0.33 &  4.69$\times10^{1}$&  4.69$\times10^{1}$&  5.66$\times10^{3}$  & 
{\bf NGC 6334}{\it(6334c)} \\
cl~237 & 17$^h$19$^m$16.5$^s$ & -36$^\mathrm{\circ}$ 2' 5.6'' &  0.14 &   0.80
 & 0.33 &  4.91$\times10^{1}$&  4.91$\times10^{1}$&  5.92$\times10^{3}$  & 
{\bf NGC 6334}{\it(6334b)} \\
cl~240 & 17$^h$20$^m$57.9$^s$ & -35$^\mathrm{\circ}$49'57.0'' &  0.14 &   0.46
 & 0.24 &  2.83$\times10^{1}$&  2.83$\times10^{1}$&  8.87$\times10^{3}$  & 
{\bf NGC 6334}{\it(6334a)}{\it(6334b)} \\
cl~241 & 17$^h$22$^m$ 7.6$^s$ & -35$^\mathrm{\circ}$25'29.6'' &  0.14 &   0.41
 & 0.24 &  2.52$\times10^{1}$&  2.52$\times10^{1}$&  7.90$\times10^{3}$  & 
{\bf NGC 6334}{\it(6334b)} \\
cl~242 & 17$^h$18$^m$17.6$^s$ & -36$^\mathrm{\circ}$ 9'47.2'' &  0.14 &   0.34
 & 0.21 &  2.12$\times10^{1}$&  2.12$\times10^{1}$&  9.94$\times10^{3}$  & 
{\bf NGC 6334}{\it(6334c)} \\
cl~243 & 17$^h$19$^m$56.7$^s$ & -35$^\mathrm{\circ}$53'42.0'' &  0.14 &   0.41
 & 0.24 &  2.53$\times10^{1}$&  2.53$\times10^{1}$&  7.94$\times10^{3}$  & 
{\bf NGC 6334}{\it(6334a)}{\it(6334b)} \\
cl~{\bf245} & 17$^h$20$^m$19.7$^s$ & -35$^\mathrm{\circ}$48'54.0'' &  0.14 & 
  0.49 & 0.27 &  3.03$\times10^{1}$&  1.25$\times10^{1}$&  2.75$\times10^{3}$
  & {\bf NGC 6334}{\it(6334a)}{\it(6334b)} \\
cl~246 & 17$^h$19$^m$51.4$^s$ & -36$^\mathrm{\circ}$ 2'30.1'' &  0.14 &   0.62
 & 0.30 &  3.80$\times10^{1}$&  3.80$\times10^{1}$&  6.11$\times10^{3}$  & 
{\bf NGC 6334}{\it(6334b)} \\
cl~247 & 17$^h$17$^m$32.8$^s$ & -36$^\mathrm{\circ}$26' 8.5'' &  0.13 &   0.66
 & 0.30 &  4.05$\times10^{1}$&  4.05$\times10^{1}$&  6.49$\times10^{3}$  & 
{\bf NGC 6334}{\it(6334c)} \\
cl~249 & 17$^h$22$^m$19.3$^s$ & -35$^\mathrm{\circ}$21'20.9'' &  0.13 &   1.42
 & 0.45 &  8.76$\times10^{1}$&  8.76$\times10^{1}$&  4.17$\times10^{3}$  & 
{\bf NGC 6334}{\it(6334b)} \\
cl~251 & 17$^h$23$^m$36.5$^s$ & -34$^\mathrm{\circ}$51'45.4'' &  0.13 &   0.32
 & 0.21 &  2.00$\times10^{1}$&  2.00$\times10^{1}$&  9.36$\times10^{3}$  & 
{\bf NGC 6334}{\it(6334c)} \\
cl~252 & 17$^h$19$^m$40.2$^s$ & -36$^\mathrm{\circ}$ 3' 9.7'' &  0.13 &   0.39
 & 0.24 &  2.38$\times10^{1}$&  2.38$\times10^{1}$&  7.45$\times10^{3}$  & 
{\bf NGC 6334}{\it(6334b)} \\
cl~253 & 17$^h$20$^m$50.1$^s$ & -35$^\mathrm{\circ}$56'37.3'' &  0.13 &   0.43
 & 0.24 &  2.64$\times10^{1}$&  2.64$\times10^{1}$&  8.27$\times10^{3}$  & 
{\bf NGC 6334}{\it(6334b)} \\
cl~256 & 17$^h$20$^m$53.4$^s$ & -36$^\mathrm{\circ}$ 1'57.4'' &  0.13 &   0.45
 & 0.27 &  2.80$\times10^{1}$&  2.80$\times10^{1}$&  6.16$\times10^{3}$  & 
{\bf NGC 6334}{\it(6334c)} \\
cl~259 & 17$^h$21$^m$ 6.1$^s$ & -35$^\mathrm{\circ}$23'33.0'' &  0.13 &   0.09
 & 0.12 &  5.56$ $&  5.56$ $&  1.39$\times10^{4}$  & 
{\bf NGC 6334}{\it(6334c)} \\
cl~260 & 17$^h$18$^m$ 9.4$^s$ & -36$^\mathrm{\circ}$20'19.0'' &  0.13 &   0.72
 & 0.33 &  4.43$\times10^{1}$&  4.43$\times10^{1}$&  5.34$\times10^{3}$  & 
{\bf NGC 6334}{\it(6334c)} \\
cl~261 & 17$^h$21$^m$17.7$^s$ & -35$^\mathrm{\circ}$51'56.5'' &  0.13 &   0.38
 & 0.24 &  2.33$\times10^{1}$&  2.33$\times10^{1}$&  7.30$\times10^{3}$  & 
{\bf NGC 6334}{\it(6334b)} \\
cl~262 & 17$^h$20$^m$35.6$^s$ & -36$^\mathrm{\circ}$ 3' 1.8'' &  0.13 &   0.19
 & 0.18 &  1.19$\times10^{1}$&  1.19$\times10^{1}$&  8.81$\times10^{3}$  & 
{\bf NGC 6334}{\it(6334b)} \\
cl~264 & 17$^h$19$^m$33.5$^s$ & -36$^\mathrm{\circ}$24'37.8'' &  0.12 &   0.21
 & 0.18 &  1.28$\times10^{1}$&  1.28$\times10^{1}$&  9.50$\times10^{3}$  & 
{\bf NGC 6334}{\it(6334c)} \\
cl~266 & 17$^h$22$^m$ 2.1$^s$ & -35$^\mathrm{\circ}$13' 5.9'' &  0.12 &   0.34
 & 0.24 &  2.13$\times10^{1}$&  2.13$\times10^{1}$&  6.68$\times10^{3}$  & 
{\bf NGC 6334}{\it(6334c)} \\
cl~268 & 17$^h$21$^m$12.7$^s$ & -35$^\mathrm{\circ}$30'36.7'' &  0.12 &   0.78
 & 0.36 &  4.83$\times10^{1}$&  4.83$\times10^{1}$&  4.49$\times10^{3}$  & 
{\bf NGC 6334}{\it(6334b)} \\
cl~271 & 17$^h$22$^m$25.1$^s$ & -35$^\mathrm{\circ}$16'56.3'' &  0.12 &   0.79
 & 0.36 &  4.90$\times10^{1}$&  4.90$\times10^{1}$&  4.55$\times10^{3}$  & 
{\bf NGC 6334}{\it(6334c)} \\
cl~273 & 17$^h$21$^m$25.8$^s$ & -35$^\mathrm{\circ}$31'32.2'' &  0.12 &   0.59
 & 0.30 &  3.66$\times10^{1}$&  3.66$\times10^{1}$&  5.88$\times10^{3}$  & 
{\bf NGC 6334}{\it(6334b)} \\
cl~276 & 17$^h$16$^m$57.0$^s$ & -36$^\mathrm{\circ}$25'25.3'' &  0.12 &   0.38
 & 0.24 &  2.37$\times10^{1}$&  2.37$\times10^{1}$&  7.41$\times10^{3}$  & 
{\bf NGC 6334}{\it(6334c)} \\
cl~279 & 17$^h$20$^m$ 5.3$^s$ & -36$^\mathrm{\circ}$27'10.1'' &  0.12 &   0.22
 & 0.18 &  1.36$\times10^{1}$&  1.36$\times10^{1}$&  1.01$\times10^{4}$  & 
{\bf NGC 6334}{\it(6334c)} \\
cl~282 & 17$^h$20$^m$12.6$^s$ & -36$^\mathrm{\circ}$ 7' 1.9'' &  0.11 &   0.17
 & 0.15 &  1.06$\times10^{1}$&  1.06$\times10^{1}$&  1.36$\times10^{4}$  & 
{\bf NGC 6334}{\it(6334b)} \\
cl~284 & 17$^h$22$^m$22.1$^s$ & -35$^\mathrm{\circ}$27' 4.7'' &  0.11 &   0.73
 & 0.36 &  4.51$\times10^{1}$&  4.51$\times10^{1}$&  4.19$\times10^{3}$  & 
{\bf NGC 6334}{\it(6334b)} \\
cl~286 & 17$^h$19$^m$42.9$^s$ & -35$^\mathrm{\circ}$59'49.9'' &  0.11 &   0.44
 & 0.27 &  2.74$\times10^{1}$&  2.74$\times10^{1}$&  6.02$\times10^{3}$  & 
{\bf NGC 6334}{\it(6334b)} \\
cl~287 & 17$^h$23$^m$37.8$^s$ & -34$^\mathrm{\circ}$52'57.4'' &  0.11 &   0.27
 & 0.21 &  1.67$\times10^{1}$&  1.67$\times10^{1}$&  7.80$\times10^{3}$  & 
{\bf NGC 6334}{\it(6334c)} \\
cl~290 & 17$^h$20$^m$ 7.3$^s$ & -36$^\mathrm{\circ}$ 7'34.0'' &  0.11 &   0.18
 & 0.18 &  1.14$\times10^{1}$&  1.14$\times10^{1}$&  8.44$\times10^{3}$  & 
{\bf NGC 6334}{\it(6334b)} \\
cl~292 & 17$^h$23$^m$48.8$^s$ & -34$^\mathrm{\circ}$52'16.0'' &  0.11 &   0.42
 & 0.24 &  2.58$\times10^{1}$&  2.58$\times10^{1}$&  8.09$\times10^{3}$  & 
{\bf NGC 6334}{\it(6334c)} \\
cl~293 & 17$^h$22$^m$13.1$^s$ & -35$^\mathrm{\circ}$35' 5.3'' &  0.11 &   0.14
 & 0.15 &  8.40$ $&  8.40$ $&  1.08$\times10^{4}$  & 
{\bf NGC 6334}{\it(6334b)} \\
cl~294 & 17$^h$23$^m$ 9.7$^s$ & -34$^\mathrm{\circ}$49'32.2'' &  0.11 &   0.51
 & 0.27 &  3.14$\times10^{1}$&  3.14$\times10^{1}$&  6.92$\times10^{3}$  & 
{\bf NGC 6334}{\it(6334c)} \\
cl~295 & 17$^h$22$^m$17.3$^s$ & -35$^\mathrm{\circ}$19'53.0'' &  0.11 &   1.43
 & 0.47 &  8.86$\times10^{1}$&  8.86$\times10^{1}$&  3.47$\times10^{3}$  & 
{\bf NGC 6334}{\it(6334b)} \\
cl~300 & 17$^h$17$^m$44.2$^s$ & -36$^\mathrm{\circ}$22'33.2'' &  0.11 &   0.19
 & 0.18 &  1.20$\times10^{1}$&  1.20$\times10^{1}$&  8.90$\times10^{3}$  & 
{\bf NGC 6334}{\it(6334c)} \\
cl~303 & 17$^h$17$^m$48.2$^s$ & -36$^\mathrm{\circ}$22'17.4'' &  0.11 &   0.24
 & 0.21 &  1.49$\times10^{1}$&  1.49$\times10^{1}$&  6.99$\times10^{3}$  & 
{\bf NGC 6334}{\it(6334c)} \\
cl~307 & 17$^h$23$^m$ 3.4$^s$ & -35$^\mathrm{\circ}$12'21.2'' &  0.11 &   0.14
 & 0.15 &  8.71$ $&  8.71$ $&  1.12$\times10^{4}$  & 
{\bf NGC 6334}{\it(6334c)} \\
cl~308 & 17$^h$23$^m$ 5.8$^s$ & -35$^\mathrm{\circ}$ 5'33.0'' &  0.11 &   0.36
 & 0.24 &  2.24$\times10^{1}$&  2.24$\times10^{1}$&  7.01$\times10^{3}$  & 
{\bf NGC 6334}{\it(6334c)} \\
cl~309 & 17$^h$18$^m$16.1$^s$ & -36$^\mathrm{\circ}$18'35.3'' &  0.11 &   0.88
 & 0.39 &  5.42$\times10^{1}$&  5.42$\times10^{1}$&  3.96$\times10^{3}$  & 
{\bf NGC 6334}{\it(6334c)} \\
cl~311 & 17$^h$22$^m$35.2$^s$ & -35$^\mathrm{\circ}$28'55.6'' &  0.11 &   0.21
 & 0.18 &  1.28$\times10^{1}$&  1.28$\times10^{1}$&  9.50$\times10^{3}$  & 
{\bf NGC 6334}{\it(6334b)} \\
cl~318 & 17$^h$17$^m$15.8$^s$ & -35$^\mathrm{\circ}$59'51.0'' &  0.10 &   0.06
 & 0.09 &  3.95$ $&  3.95$ $&  2.35$\times10^{4}$  & 
{\bf NGC 6334}{\it(6334c)} \\
cl~319 & 17$^h$16$^m$58.0$^s$ & -36$^\mathrm{\circ}$17'49.6'' &  0.10 &   0.87
 & 0.36 &  5.36$\times10^{1}$&  5.36$\times10^{1}$&  4.98$\times10^{3}$  & 
{\bf NGC 6334}{\it(6334c)} \\
cl~326 & 17$^h$21$^m$34.5$^s$ & -35$^\mathrm{\circ}$37'39.7'' &  0.10 &   0.12
 & 0.15 &  7.16$ $&  7.16$ $&  9.20$\times10^{3}$  & 
{\bf NGC 6334}{\it(6334b)} \\
cl~329 & 17$^h$20$^m$24.2$^s$ & -35$^\mathrm{\circ}$29'25.8'' &  0.10 &   0.17
 & 0.18 &  1.06$\times10^{1}$&  1.06$\times10^{1}$&  7.89$\times10^{3}$  & 
{\bf NGC 6334}{\it(6334c)} \\
cl~336 & 17$^h$19$^m$42.8$^s$ & -36$^\mathrm{\circ}$ 3' 1.8'' &  0.10 &   0.28
 & 0.21 &  1.73$\times10^{1}$&  1.73$\times10^{1}$&  8.09$\times10^{3}$  & 
{\bf NGC 6334}{\it(6334b)} \\
cl~337 & 17$^h$19$^m$42.8$^s$ & -36$^\mathrm{\circ}$23'25.8'' &  0.10 &   0.29
 & 0.21 &  1.80$\times10^{1}$&  1.80$\times10^{1}$&  8.41$\times10^{3}$  & 
{\bf NGC 6334}{\it(6334c)} \\
cl~340 & 17$^h$20$^m$ 8.6$^s$ & -36$^\mathrm{\circ}$ 6'38.2'' &  0.10 &   0.13
 & 0.15 &  8.21$ $&  8.21$ $&  1.05$\times10^{4}$  & 
{\bf NGC 6334}{\it(6334b)} \\
cl~342 & 17$^h$21$^m$25.9$^s$ & -35$^\mathrm{\circ}$32'36.2'' &  0.10 &   0.17
 & 0.15 &  1.05$\times10^{1}$&  1.05$\times10^{1}$&  1.35$\times10^{4}$  & 
{\bf NGC 6334}{\it(6334b)} \\
cl~343 & 17$^h$22$^m$20.4$^s$ & -35$^\mathrm{\circ}$35'12.8'' &  0.10 &   0.20
 & 0.18 &  1.27$\times10^{1}$&  1.27$\times10^{1}$&  9.41$\times10^{3}$  & 
{\bf NGC 6334}{\it(6334b)} \\
cl~344 & 17$^h$20$^m$25.0$^s$ & -35$^\mathrm{\circ}$41'10.0'' &  0.10 &   0.21
 & 0.18 &  1.28$\times10^{1}$&  1.28$\times10^{1}$&  9.54$\times10^{3}$  & 
{\bf NGC 6334}{\it(6334b)} \\

\enddata
\tablenotetext{a}{$M_\mathrm{1.2mm}$ is the computed clump mass 
from all the emssion at 1.2~mm and assuming a
temperature of 17~K, while $M$ considers
the two-temperature clump ensemble and the correction by free-free emission.}
\tablenotetext{b}{The average number density was computed using 
$R_\mathrm{eff}$ and $M$ and assuming spherical clumps with a mean
molecular weight of $\mu=2.3$. }
\tablecomments{The clump numbers highlighted in the first column
are those which are considered to be significantly warmer. Clump
210 has an undetermined mass after subtracting the free-free
emission at 1.2~mm}
\end{deluxetable*}


\begin{thebibliography}{}
\bibitem [Altenhoff et al.(1961)]{alten61}
Altenhoff, W., Mezger, P.G., Wendker, H., \& Westerhout, G., 1961, Veroffentl.\ Sternwarte, 59, 48 
\bibitem[Alves et al.(2007)]{alv07} Alves, J., Lombardi, M., 
\& Lada, C.~J.\ 2007, \aap, 462, L17 
\bibitem[Ballesteros-Paredes et al.(1999)]{ball99}
Ballesteros-Paredes, J., V{\'a}zquez-Semadeni, E., \& Scalo, J.\
1999, \apj, 515, 286
\bibitem[Beckert et al.(2000)]{bec00} Beckert, T., Duschl,
W.~J., \& Mezger, P.~G.\ 2000, \aap, 356, 1149
\bibitem[Bertoldi \& McKee(1992)]{bert92} Bertoldi, F., \&
McKee, C.~F.\ 1992, \apj, 395, 140
\bibitem[Beuther \& Schilke(2004)]{beu04}Beuther, H. \& Schilke, P. 2004,
Science, 303, 1167
\bibitem[Beuther et al.(2006)]{beu06} Beuther, H.,
Churchwell, E.~B., McKee, C.~F., \& Tan, J.~C.\ 2006, ArXiv Astrophysics
e-prints, arXiv:astro-ph/0602012
\bibitem[Bonnell \& Davies(1998)]{bonn98} Bonnell, I.~A., \& Davies, M.~B.\ 1998, \mnras, 295, 691
\bibitem[Blitz(1993)]{bli93} Blitz, L. 1993, Giant Molecular Clouds,
in Protostars and Planets III, eds. E.H. Levy \& J.I. Lunine
(Tucson: Univ. of Arizona Press)
\bibitem[Brooks et al.(2005)]{brook05} Brooks, K.~J., Garay, 
G., Nielbock, M., Smith, N., \& Cox, P.\ 2005, \apj, 634, 436 
\bibitem[Burton et al.(2000)]{burt00} Burton, M.G., Ashley, M.C.B.,
    Marks, R.D., Schinckel, A.E., Storey, J.W.V.,
    Fowler, A., Merril, M., Sharp, N., Gatley, I., Harper, D.A.,
    Loewenstein, R.F., Mrozek, F., Jackson, J.M., \& Kraemer, K.E.
    2000 \apj,542, 359
\bibitem[Carral et al.(2002)]{carr02} Carral, P.,
    Kurtz, S.E., Rodr\'iguez, L.F., Menten, K.,
    Cant\'o, J., \&
    Arceo, R.  1978, \aj, 123, 2574
\bibitem[Chini, Kr\"ugel \& Wargau (1987)]{chi87} Chini, R.,
    Kr\"ugel, E. \& Wargau, W. 1987, A\&A, 181, 378
\bibitem[Clark \& Bonnell(2006)]{cla06} Clark, P.~C., \& Bonnell, I.~A.\ 2006, \mnras, 368, 1787
\bibitem[Clark \& Bonnell(2005)]{cla05} Clark, P.~C., \& Bonnell, I.~A.\ 2005, \mnras, 361, 2
\bibitem[Evans(1999)]{eva99} Evans, N.~J., II 1999, \araa,
37, 311
 \bibitem[Fa{\'u}ndez et al.(2004)]{fau04} Fa{\'u}ndez, S.,
Bronfman, L., Garay, G., Chini, R., Nyman, L.-{\AA}., \& May, J.\
2004, \aap, 426, 97
\bibitem[Garay et al.(2002)]{gar02} Garay, G., Brooks, K.J., Mardones, D.,
Norris,R.P., \& Burton, M.G. 2002, \apj, 579, 678
\bibitem[Gezari(1982)]{gez82} Gezari, D.Y. 1982, \apj, 259, L29
\bibitem[G\'omez et al.(1993)]{gom93} G\'omez, M.,Hartmann, S.,
    Kenyon, S.J.,\& Hewett,R.  1993, \aj, 105, 5
    \bibitem[Hunter et al.(2006)]{hun06} Hunter, T.~R., Brogan, 
C.~L., Megeath, S.~T., Menten, K.~M., Beuther, H., \& Thorwirth, S.\ 2006, \apj, 649, 888
\bibitem[Jackson \& Kraemer(1999)]{jack99} Jackson, J.M. \& Kraemer, K. E., 1999, \apj, 512, 260
\bibitem[Johnstone et al.(2000a)]{john00a} Johnstone, D.,
Wilson, C.~D., Moriarty-Schieven, G., Giannakopoulou-Creighton, J., \& Gregersen, E.\ 2000a, \apjs, 131, 505
\bibitem[Johnstone et al.(2000b)]{john00b} Johnstone, D., Wilson, C.D.,
Moriarty-Schieven, G., Joncas, G., Smith, G., Gregersen, E. \& Fich, M.
2000b, \apj, 545, 327
\bibitem[Johnstone et al.(2001)]{john01} Johnstone, D.,Fich, M.
Mitchell, G.F., \& Moriarty-Schieven,
2001, \apj, 559, 307
\bibitem[Johnstone et al.(2006)]{john06} Johnstone, D.,
Matthews, H., \& Mitchell, G.~F.\ 2006, \apj, 639, 259
\bibitem[Kerton et al.(2001)]{ker01} Kerton, C.R., Martin, P.G.,,
Johnstone, D., \& Ballantyne, D.R. 2001, \apj, 552, 601
\bibitem[Kitsionas et al.(1998)]{kit98} Kitsionas, S.,
Gladwin, P.~P., \& Whitworth, A.~P.\ 1998, ASP Conf.~Ser.~132: Star
Formation with the Infrared Space Observatory, 132, 434
\bibitem[Klessen(2001)]{kless01} Klessen, R.~S.\ 2001, \apj,556, 837
\bibitem[Klessen \& Burkert(2000)]{kle00} Klessen, R.~S., \& Burkert, A.\ 2000, \apjs, 128, 287
\bibitem[Kramer et al.(1998)]{kram98} Kramer, C.,
  Stutzki, J., R\"orig, R.,\& Corneliussen, U. 1998, A\&A, 329, 249
\bibitem[Kraemer \& Jackson(1999)]{krae99b} Kraemer, K.~E., \& Jackson, J.~M.\ 1999, \apjs, 124, 439
\bibitem[Kraemer et al.(2000)]{krae00} Kraemer, K.E.,
  Jackson, J.M. 1999, Lane, A.P. \& Paglione, T.A.D. \apj, 542, 946
\bibitem[Kroupa(2001)]{kro01} Kroupa, P.\ 2001, \mnras, 322,
231
\bibitem[Lada \& Lada(2003)]{lad03} Lada, C.~J., \& Lada,
E.~A.\ 2003, \araa, 41, 57
\bibitem[Lada(1999)]{elad99} Lada, E.~A.\ 1999, NATO ASIC
Proc.~540: {\it The Origin of Stars and Planetary Systems}, C.J. Lada,
N.D. Kylafis, Eds. (Kluwer, Dotrecht, Netherlands, 200), p 441
\bibitem[Larson(1995)]{lar95} Larson, R. 1995, \mnras, 272, 213
\bibitem[Larson(2003)]{lar03} Larson, R.~B.\ 2003, ASP
Conf.~Ser.~287: {\it Galactic Star Formation Across the Stellar Mass Spectrum},
J.M. De Buizer abd N.S. van der Blick, (eds), 287, 65
\bibitem[Li et al.(2006)]{li06} Li, D., Velusamy, T., 
Goldsmith, P.~F., \& Langer, W.~D.\ 2006, ArXiv Astrophysics e-prints, 
arXiv:astro-ph/0610634 
\bibitem[McBreen et al.(1979)]{mcbree79} McBreen,B., Fazio, G.G.,
    Stier, M., \& Wright, E.L. 1979, \apj, 232, L183
\bibitem[McCutcheon et al.(2000)]{mccut00} McCutcheon, W.H., Sandell, G.
    Matthews, H.E., Kuiper, T.B.H., Sutton, E.C., Danchi, W.C. \& Sato, T.
    2000, \mnras, 316, 152
 \bibitem[Menzel \& Pekeris(1935)]{men35} Menzel, D.~H., \&
Pekeris, C.~L.\ 1935, \mnras, 96, 77
\bibitem[Mookerjea et al.(2005)]{mook05} Mookerjea, B., Kramer, C.,
Nielbock, M., \& Nyman, L.-\AA.  2005, A\&A, 426, 119
\bibitem[Moran et al.(1990)]{mor90} Moran, J.~M., Greene, B., Rodriguez, L.~F., \& Backer, D.~C.\ 1990, \apj, 348, 147
\bibitem[Motte et al.(1998)]{mott98} Motte, F., Andre, P., \&
Neri, R.\ 1998, \aap, 336, 150
\bibitem[Motte et al.(2003)]{mott03} Motte, F., Schilke, P., \& Lis, D.C., 2003, \apj, 582, 277	
\bibitem[Mu\~noz(2006)]{mun06a} Mu\~noz, D.J.
\ 2006, M.Sc.~Thesis, Universidad de Chile
\bibitem[Nakamura \& Li(2005)]{nak05} Nakamura, F., \& Li,
Z.-Y.\ 2005, \apj, 631, 411
\bibitem[Neckel(1978)]{neck78} Neckel, T. 1978, A\&A, 69, 51
\bibitem[Nozawa et al.(1991)]{noz91}Nozawa, S., Mizuno, A., Teshima, Y.,
  Ogawa, H. \& Fukui, Y. 1991, \apjs, 77,647
\bibitem[Onishi et al. (1996)]{oni96} Onishi, t., Mizuno, A., Kawamura, A., Ogawa, H., \& Fukui, Y. 1996, \apj, 465, 815
\bibitem[Ossenkopf \& Henning(1994)]{oss94} Ossenkopf, V., \& Henning, T.\ 1994, \aap, 291, 943
\bibitem[Padoan \& Nordlund(2002)]{pado02} Padoan, P., \& Nordlund, \AA\  2002,  \apj, 576, 870
\bibitem[Plume et al.(1997)]{plum97} Plume, R., Jaffe, D.~T., Evans, N.~J., II, Martin-Pintado, J., \& Gomez-Gonzalez, J.\ 1997, \apj, 476, 730
\bibitem[Pudritz(2002)]{pud02} Pudritz, R.~E.\ 2002, Science,
295, 68
\bibitem[Reid \& Wilson(2005)]{rei05} Reid, M.~A., \& Wilson,
C.~D.\ 2005, \apj, 625, 891
\bibitem[Reid \& Wilson(2006a)]{rei06a} Reid, M.~A., \& Wilson, 
C.~D.\ 2006a, \apj, 644, 990 
\bibitem[Reid \& Wilson(2006b)]{rei06b} Reid, M.~A., \& Wilson, 
C.~D.\ 2006b, \apj, 650, 970
\bibitem[Rodr\'iguez, Cant\'o \& Moran(1982)]{rod82} Rodr\'iguez, L.F.,
Cant\'o, J., \& Moran, J.M.  1982, \apj, 255, 103
\bibitem[Rosolowsky(2005)]{roso05} Rosolowsky, E.\ 2005, \pasp, 117, 1403
\bibitem[Rybicki \& Lightman(1979)]{ryb79} Rybicki, G.~B., \&
Lightman, A.~P.\ 1979, {\it Radiative Processes in Astrophysics},
New York, Wiley-Interscience, 1979.~393 p.,
\bibitem[Salpeter(1955)]{sal55} Salpeter, E.~E.\ 1955, \apj,
121, 161
\bibitem[Sandell(2000)]{san00} Sandell, G.  2000, A\&A, 358,242
\bibitem[Sanders et al.(1985)]{san85} Sanders, D.~B.,
Scoville, N.~Z., \& Solomon, P.~M.\ 1985, \apj, 289, 373
\bibitem[Scalo(1998)]{sca98} Scalo, J.\ 1998, ASP
Conf.~Ser.~142: {\it The Stellar Initial Mass Function} (38th Herstmonceux
Conference), G.~Gilmore  and D.~Howell (eds),142, 201
\bibitem[Schneider \& Brooks(2004)]{schnei04} Schneider, N., \& 
Brooks, K.\ 2004, Publications of the Astronomical Society of Australia, 
21, 290 
\bibitem[Silverman(1986)]{silv86} Silverman, B.~W.\ 1986,
Monographs on Statistics and Applied Probability, London: Chapman and Hall,
1986
\bibitem[Solomon et al.(1987)]{solo87} Solomon, P.~M., Rivolo,
A.~R., Barrett, J., \& Yahil, A.\ 1987, \apj, 319, 730
\bibitem[Sommerfeld(1953)]{som53}Sommerfeld, A.J.F., 
{\it Atombau und Spektrallinien}, Vol.2. Ungar, New York, 1953.
\bibitem[Straw \& Hyland(1989)]{stra89b} Straw, S.~M., \& Hyland, A.~R.\ 1989, \apj, 340, 318 
\bibitem[Straw et al.(1989)]{stra89a} Straw, S.~M., Hyland, 
A.~R., \& McGregor, P.~J.\ 1989, \apjs, 69, 99 
\bibitem[Stutzki \& Gusten(1990)]{stut90} Stutzki, J., \& G\"usten, R.
1990, \apj, 356, 513
\bibitem[Testi \& Sargent(1998)]{tes98} Testi, L., \&
Sargent, A.~I.\ 1998, \apjl, 508, L91
\bibitem[Tilley \& Pudritz(2005)]{till05} Tilley, D.~A., \&
Pudritz, R.~E.\ 2005, ArXiv Astrophysics e-prints,
arXiv:astro-ph/0508562
\bibitem[Tothill et al.(2002)]{tot02} Tothill,N.F.H., White, G.J.,
Matthews, H.E., McCutcheon, W.H., McCaughrean, M.J., \& Kenworthy, M.A.
2002, \apj, 580, 285
\bibitem[V{\'a}zquez-Semadeni et al.(2005)]{vaz05}
V{\'a}zquez-Semadeni, E., Kim, J., Shadmehri, M., \&
Ballesteros-Paredes, J.\ 2005, \apj, 618, 344
\bibitem[Ward-Thompson et al.(2006)]{war06} Ward-Thompson,
D., Andre, P., Crutcher, R., Johnstone, D., Onishi, T., \& Wilson, C.\
2006, ArXiv Astrophysics e-prints, arXiv:astro-ph/0603474
\bibitem[Williams, Blitz \& McKee(2000)]{will00} Williams, J.P., Blitz,L.
\& McKee, C.F. 2000, in Protostars and Planets IV, ed. V. Mannings, A.P. Boss,
\& S.S. Russell (Tucson: Univ. of Arizona), 97
\bibitem[Williams, de Geus \& Blitz(1994)]{will94} Williams, J.P., de Geus
,E.J., \& Blitz, 1994, \apj, 428, 693
\bibitem[Yonekura et al.(2005)]{yone05} Yonekura, Y., Asayama,
S., Kimura, K., Ogawa, H., Kanai, Y., Yamaguchi, N., Barnes,
P.~J., \& Fukui, Y.\ 2005, \apj, 634, 476
\end{thebibliography}
\end{document}